\documentclass{aa}
\usepackage{graphicx,psfig}
\usepackage{natbib}
\usepackage{txfonts}
\def\be{\begin{equation}}
\def\ee{\end{equation}}
\begin{document}
\title{Planet Signatures in Collisionally Active Debris Discs: scattered light images}

\author{P. Thebault\inst{1}, Q. Kral\inst{1}, S. Ertel\inst{2}}
\institute{LESIA-Observatoire de Paris, CNRS, UPMC Univ. Paris 06, Univ. Paris-Diderot, France
\and
Universit\'e Joseph Fourier/CNRS, LAOG, UMR5571, Grenoble, France
}

\offprints{P. Thebault} \mail{philippe.thebault@obspm.fr}
\date{Received; accepted} \titlerunning{Planet signatures in debris discs}
\authorrunning{Thebault et al.}

\abstract
%
% Context: -------------------
{Planet perturbations have been often invoked as a potential explanation for many spatial structures that have been imaged in debris discs. So far this issue has been mostly investigated with pure N-body numerical models, which neglect the crucial effect collisions within the disc can have on the disc's response to dynamical perturbations.
}
%
% Aims: ------------------------
{We numerically investigate how the coupled effect of collisions and radiation pressure can affect the formation and survival of radial and azimutal structures in a disc perturbed by a planet. We consider two different set-ups: a planet embedded within an extended disc and a planet exterior to an inner debris ring. One important issue we want to address is under which conditions a planet's signature can be observable in a collisionally active disc.
}
% Methods:
{We use the DyCoSS code of Thebault(2012), which is designed to investigate the structure of perturbed debris discs at dynamical and collisional steady-state, and derive synthetic images of the system in scattered light. The planet's mass and orbit, as well as the disc's collisional activity (parameterized by its average vertical optical depth $\tau_0$) are explored as free parameters.}
% Results:
{We find that collisions always significantly damp planet-induced spatial structures. For the case of an embedded planet, the planet's signature, mostly a density gap around its radial position, should remain detectable in head-on images if $M_{planet}\geq M_{Saturn}$. If the system is seen edge-on, however, inferring the presence of the planet is much more difficult, as only weak asymmetries remain in a collisionally active disc, although some planet-induced signatures might be observable under very favourable conditions. 

For the case of an inner ring and an external planet, planetary perturbations cannot prevent collision-produced small fragments from populating the regions beyond the ring. The radial luminosity profile exterior to the ring is in most cases close to the one it should have in the absence of the external planet. The most significant signature left by a Jovian planet on a circular orbit are precessing azimutal structures that can be used to indirectly infer its presence. For a planet on an eccentric orbit, we show that the ring becomes elliptic and that the well known pericentre glow effect is visible despite of collisions and radiation pressure, but that detecting such features in observed discs is not an unambiguous indicator of the presence of an outer planet.
}
 {}

\maketitle

\section{Introduction} \label{intro}

Most imaged dusty debris discs\footnote{30 as of July 2012, see http://circumstellardisks.org/} display pronounced radial and azimutal structures, which are the sign that these systems are dynamically active and that "something" is shaping them. Depending on the type of spatial features, i.e., two-sided asymmetries, spirals, warps, clumps or rings \citep[e.g][]{kala05,goli06,schn09}, several possible explanations have been investigated: transient violent events \citep{keny05,grig07}, coupling to gas drag \citep{take01}, companion star perturbations \citep{auge04,theb10,theb12} or dynamically cold discs \citep{theb08}.
However, the most commonly proposed scenario for the vast majority of theses structures is the presence of (usually unseen) perturbing planets. The dynamical effect of a(several) planet(s) on a disc has been investigated in numerous studies, usually numerical investigations based on deterministic N-body codes. These investigations have lead to several important results, notably that dust can easily be captured in mean motion resonances with inner planets that are migrating, thus creating, depending on resonance order and planet orbits, pronounced azimutal and radial overdensities \citep[e.g.][]{kuch03,wyat06,rech08}. Likewise, the chaotic region surrounding an embedded planet can efficiently truncate a disc, inducing sharp inner or outer edges and ring-like structures \citep{wisd80,must12}. Embedded planets can also trigger transient spiral structures that can often be long-lived enough to be observed in extended debris discs \citep{wyat05}. Finally, inclined planets have been found to be a possible cause for warped discs such as in $\beta$-Pictoris \citep{moui97,auge01,daws11,chau12}.

However, caution is required when trying to directly match N-body code results to imaged discs. Indeed, most resolved images are obtained in scattered light, at wavelengths for which the flux is dominated by the smallest dust grains on bound orbits, which are strongly affected by stellar radiation pressure \citep{theb07}. While including the effect of radiation pressure in N-body codes is not difficult in itself, usually by correcting the stellar gravity by a factor $(1-\beta)^{0.5}$ (where $\beta$ is the ratio between radiation pressure and stellar gravity), the fact that it is a size-dependent effect requires to have an idea of the \emph{size distribution} in the disc. The problem is that the size distribution is imposed by the $collisional$ evolution of the system, and that collisions are not taken into account in pure N-body codes. To circumvent this problem, an idea would be to perform series of dynamics+radiation-pressure N-body runs, each for a given particle size, and then recombine them with a weight given by theoretical collisional-equilibrium size distributions; for example the classical differential distribution in -3.5 of \citet{dohn69} \citep{moro02,quil02,wyat06,erte12}. But the implicit assumption behind this procedure, i.e., that size distributions are not affected by spatial structures or dynamical perturbations, has been proven to be faulty. As a matter of fact, \citet{stru06} and \citet{theb08} have shown that, even in an \emph{unperturbed} system consisting of a narrow parent body ring, collisional evolution naturally creates a disc-integrated overabundance of small grains close to the blow-out limit, and thus a strong departure from a "classical" collisional equilibrium distribution. Size distributions in perturbed systems could depart even more from standard power-laws, with potentially strong spatial variations \footnote{We note that there is also observational support for wavelength-dependent (and therefore grain size-dependent) structures in debris discs \citep[e.g.][]{su05,wilne11}}. It is thus very difficult to know in advance the relative contributions of different grain populations to the system's luminosity.

The non-inclusion of collisions in N-body studies has other problematic consequences. An important issue is that of the \emph{timescales}: if the typical collision times are shorter than the dynamical timescales required for spatial structures to form, then these structures' development can be affected or even hindered. This is why pure dynamical modelling, such as the recent investigation by \citet{erte12} unavoidably overestimate the level of spatial structures in dusty discs, especially dense ones. Another issue is the unavoidable feedback of the collisions on the dynamics, as they dissipate energy and produce fragments on new orbits. Last but not least, the steady production of small collisional fragments, coupled to the effect of radiation pressure on them, can inject matter in dynamically unstable regions before planets can efficiently remove them.

Including collisions, especially fragmenting ones as in debris discs, into a N-body scheme has proven a very arduous task. "Brute force" methods, where test particles are effectively broken into fragments that are then dynamically evolved \citep{beau90}, are in principle the most reliable ones, but lead to an exponential increase of particle numbers that is very quickly unmanageable. Another alternative is to run pure N-body runs to get an idea of the dynamical effect of planets, and then use the results, in the form of averaged laws for impact probabilities and collision velocities, in classical particle-in-a-box collisional codes \citep[][ and references therein]{keny08}. This approach remains however limited to 1-D spatial resolution and cannot study the formation of fine structures. The most promising approach is probably a hybrid model, where each particle of the N-body code is a "super particle" (SP) standing for a cloud of real bodies sharing a common physical size, whose mutual collisions are then treated with a particle-in-a-box scheme. The first versions of such a model \citep{grig07} had a progressive increase of the number of SPs and where limited to short timescales. However, the recent "LIDT" code developed by \citet{char12} has solved this problem but is so far operational only for the simpler case of low velocity (i.e., mostly accreting) collisions in proto-planetary discs.

So far, only two models have managed to incorporate, to some extent, collisional effects into N-body codes in the debris-disc case of high-velocity fragmenting impacts. Both codes are comparable in terms of the level of collisional effects they include: they do not fully couple collisions and dynamics in a self-consistent way but are designed to study systems perturbed by $one$ perturber, once a dynamical and collisional \emph{steady state} has been reached.
The first one is the Collisional Grooming Algorithm (CGA) of \citet{star09}. Its principle is to first perform "seed runs" of collisionless particles, from which streams of successive particle positions are recorded to construct density maps. These maps are then used to derive collision destruction probabilities in new seed runs, from which new density maps are derived, etc. The process is iterated until convergence is reached. This code is limited (so far) to the case of one internal perturber on a circular orbit\footnote{an improved, more versatile version of CGA handling eccentric perturbers is under development (Stark, personal communication)}, but has given impressive results for the Kuiper Belt \citep{kuch10}. The second code is the "DyCoSS" (for "Dynamics and Collisions at Steady State") algorithm of \citet[][ hereafter TBO12]{theb12}. Although being significantly less user-friendly in its use and lacking the intrinsic coherence of the CGA, it is (in their respective present versions) more versatile with respect to the perturber's orbit, which can be circular or eccentric, and external to, interior to or embedded in the disc. It is also better suited to the study of short timescale processes, such as the competing effect of dynamical removal and collisional production of small, radiation-pressure-affected grains in high-density discs.

We use here an updated version of DyCoSS, which had initially been developped to study circumprimary discs in binaries (TBO12), adapted to the case of a planetary perturber. The main issue we are planning to address is how the collisional activity inside a disc affects the dynamical sculpting a planet excerts on it, in particular because of the constant flow of small collision fragments flying around the system because of radiation pressure. One crucial point is under which conditions a planet's signature can be detectable in a collisionally active disc, and if this signature can used as diagnostic of the presence of a planet.

\section{The DyCoSS code} \label{model}

In \citet{theb12}, we give a thorough description of the (then still unnamed) DyCoSS algorithm. Let us here only recall its main characteristics and the improvements that have been implemented for the planet-in-a-disc version.

\subsection{Principle}

As mentioned earlier, the basic set-up for DyCoSS is a collisionally active disc and one perturbing body. The main idea behind the code is that the grains that populate the disc at a given time $t_0$, when the perturber is at a position angle $\phi_0$ on its orbit, are the superposition of grains produced at different moments in the past.
Considering a discrete time increment $\Delta t$, then grains present at present time $t_0$ are the sum of grains that have just been produced (at $t_0$); plus grains produced at $t_0-\Delta t$, when the perturber was at position angle $\phi_{-1}$, which have not been destroyed (by collisions) or ejected (by dynamical perturbations) between $t_0-\Delta t$ and now; plus grains produced at $t_0-2\Delta t$, when the perturber was at position angle $\phi_{-2}$, which have not been destroyed or ejected between $t_0-2\Delta t$ and now, and so on... So, if we know the fate of particles released at $t_0-\Delta t$, at $t_0-2\Delta t$, etc., then a map of grain densities at $t_0$ can be reconstructed. This can be done by performing a series of separate runs, one with grains released when the perturber initally is at $\phi_0$, one for $\phi_{-1}$, one for $\phi_{-2}$, etc.. For each of these runs the positions of all surviving particles are recorded after every $\Delta t$ time interval, the number of remaining grains progressively decreasing because of dynamical ejection or collisional destruction. The surface density $\Sigma$ of the disc at $t_0$ is then obtained from 
\begin{equation}
\Sigma(t_{0}) = \sum_{i=0}^{\infty} \sigma(t_0)_{(t_0-i\cdot\Delta t)} 
\label{princip}
\end{equation}
where $\sigma(t_0)_{(t_0-i\cdot\Delta t)}$ is the surface density, at $t_0$, of the run started at $t_0-i\cdot\Delta t$.
The problem is of course that this procedure would require an almost infinite number of separate runs, one starting at each initial time $t_0-i\cdot\Delta t$. This problem can be circumvented if we assume that the system has reached a steady state. In this case the disc's profile is the same at successive passages of the perturber (here, the planet) at the same orbital phase. If for example, the planet's orbital period $t_{orb}$ is equal to $n_{orb}\,\Delta t$, then all the runs with the planet at initial position $\phi_{-i}$, $\phi_{-i-n_{orb}}$, $\phi_{-i-2n_{orb}}$, etc., do correspond to the same run, so that in practice only $n_{orb}$ separate runs are needed.

The numerical procedure is divided into 3 steps (see TBO12 for a more complete description):
\begin{itemize}
\item 1) \emph{Parent Body run}. A collisionless pure N-body run is performed, where radiation pressure effects are ignored, until $dynamical$ steady state is reached. The spatial distribution of the parent body particles is then recorded for a sample of $n_{orb}$ different orbital positions of the planet on its orbit.
\item 2) \emph{Collisional runs}. From each of these $n_{orb}$ steady-state parent-body discs, $N=2\times 10^{5}$ small grains are released following a $dN \propto s^{q}ds$ power law. The grains' evolution is then followed, taking this time into account the effect of radiation pressure. Depending on its size and the local optical depth in the disc, each grain is assigned, at each timestep, a collision destruction probability that depends on grain size, local velocity and the local geometrical vertical optical depth $\tau_{r,\theta}$. The relative spatial distribution of $\tau_{r,\theta}$ is obtained from maps of the steady-state parent body runs and its magnitude is scaled by the system's average optical depth $\tau_0$, which is treated as a free input parameter.
All particle positions are recorded at each $\Delta t=t_{orb}/{n_{orb}}$ interval. Runs are stopped once all particles have been removed by dynamical ejection or collision.
\item 3) \emph{Recombining}. For each orbital position of the perturber, the dynamical + collisional steady-state of the disc is obtained by recombining the position data stored at step (2) following the procedure given in Equ.\ref{princip}; each time assuming that $t_0$ is the time where the planet is at a given orbital position $\phi(i)_{0\leq i\leq n_{norb}}$.
\end{itemize}

Results are then displayed in the form of synthetic images and surface brightness profiles, in scattered light, assuming grey scattering for all particles. They are thus valid at any wavelength $\lambda$ where the flux is dominated by scattered light emission. Using the Debris Disc Simulator (DDS) of \citet{wolf05} for a typical debris disc located at 50\,AU from its star, we find that the corresponding wavelength domain is $\lambda \leq 8\mu$m for a low mass solar-type star (in accordance with the result displayed in Fig.7 of \citet{kuch10}) and $\lambda \leq 5\mu$m for a beta-Pic like A star.

We also implicitly assume that the smallest particles in the system, corresponding to the radiation-pressure blowout size $s_{cut}$, are bigger than $\lambda/2\pi$, so that all particles contribute to the scattered-light flux as a function of their geometrical cross section. Note that the absolute value for the cut-off size $s_{cut}$ is of no importance in our set-up. All simulation results can be scaled up or down depending on the real value of $s_{cut}$, as long as the star is luminous enough, at least $\sim 0.9M_{\odot}$, so that there \emph{is} such a cut-off size. For this smallest possible stellar mass $s_{cut} \sim 1\mu$m for compact silicates, which means that our $s \geq \lambda/2\pi$ condition is valid up to $\lambda\sim 6\mu$m in this most constraining case, thus approximately corresponding to the whole domain of scattered-light dominated luminosity.

\subsection{Improvements}

Several upgrades have been implemented with respect to the initial DyCoSS version. A first important update is in the way the collision destruction probability is estimated in the collisional runs. In TBO12, the local geometrical optical depths $\tau_{r,\theta}$, which control the particles' collisional timescales, were derived using a 2-D density map that was obtained from an average of $n_{orb}$ steady-state parent body disc profiles. This map was thus in practice azimutally averaged and only had 1-D information, an assumption that was acceptable for the external-stellar-perturber case where azimutal structures were limited. For an embedded planet, however, azimutal structures are expected to be more prominent and we must retain the azimutal information. As a consequence, we use the parent body runs to produce 100 density maps, corresponding to 100 different orbital positions\footnote{This increment is decoupled and in general much greater than $n_{orb}$} of the planet, which are then used in the collisional runs (the density map used at a given time is the one corresponding to the planet location closest to the present one).

Apart from this upgrade, the collision prescription is the same as in TBO12 and is given by Equ.1 of that paper. We follow \citet{star09} and assume that collisions between similarly-sized dust grains are fully destructive, i.e., particles are removed when suffering one such impact. This simplifying prescription has been taken because of numerical constraints, as a realistic treatment of fragmenting collisions is still impossible in an N-body code approach (see discussion in Section 1). It is of course not fully realistic, as collisions cannot fully vaporize particles on the spot and should produce clouds of smaller fragments. Note, however, that Equ.1 of TBO12 estimates the collision rate between \emph{similarly-sized} bodies and that the typical impact velocities in debris discs are expected to be high enough for such collisions to lead to catastrophic fragmentation and thus produce a largest-remaining-fragment much smaller than the impactor \citep[e.g., ][]{theb09} \footnote{This high-velocity assumption is confirmed in our runs, where the average collision velocities are of the order of $\sim 500\,$m.s$^{-1}$. This is much higher than the typical velocity, $\sim 10-50$m.s$^{-1}$ that is required to catastrophically destroy micron-sized grains \citep{theb09}} so that one can assume that a given impactor loses its identitys after such an impact. Furthermore, for the smallest grains that dominate the system's brightness in scattered light, of sizes between the radiation pressure blow-out size $s_{cut}$ and a few $s_{cut}$ \citep{theb07}, all produced collisional fragments will be smaller than $s_{cut}$ and thus very quickly removed from the system.
As such, our prescription gives a satisfying estimate of a typical particle's collisional lifetime, which is the main parameter of interest in the present study.
 
Another update is that Poynting-Robertson drag is now taken into account for the collisional runs. Finally, in order to have a better statistics for a wider range of grain sizes, we perform 2 collisional runs for each given position $\phi_{i}$ of the planet, one for small grains between $s_{cut}$ (corresponding to a $\beta$ value of 0.5), the cut-off size imposed by radiation pressure, and $5s_{cut}$, and one for big grains between $5s_{cut}$ and $40s_{cut}$. These two runs are then merged, weighting the big-grains run by a factor $(40s_{cut}^{q+1}-5s_{cut}^{q+1})/(5s_{cut}^{q+1}-s_{cut}^{q+1})$ to account for the size distribution with slope $q$.

\subsection{set-up} \label{set-up}

We consider two different set-ups. The first one is a planet embedded inside an extended debris disc, the second one is a planet exterior to an inner ring-like disc. We consider a fiducial "standard" debris disc extending from 30 to 130\,AU (the typical stellar distance at which most structures in resolved debris discs have been imaged) for the embedded planet case, and from 45 to 75\,AU for the exterior planet configuration. As for the planet, we take its semi-major axis to be $a_p =75\,$AU. 
We consider as a reference case a planet of mass $M_{p}/M_\star=\mu=2\times 10^{-3}$ on a circular orbit $e_p=0$ and a system of mean optical depth $\tau_0=2\times 10^{-3}$ corresponding to a dense debris disc like $\beta$ Pic.
Note that the absolute values of the disc and planet radial positions do not strongly affect the results and that our results are easily scalable (see Sec.\ref{Discu}). What matters here mostly are the disc's assumed geometrical optical depth $\tau_0$, which controls its collisional evolution, and the planet's mass $M_p$ and eccentricity $e_p$, which determine its efficiency in perturbing the disc and ejecting unstable particles. All these 3 parameters are explored as free parameters.

All set-up parameters are summarized in Tab.\ref{init}.

\begin{table}
\begin{minipage}{\columnwidth}
\caption[]{Set-up for the nominal case runs. Parameters denoted by a * are explored as free parameters}
\renewcommand{\footnoterule}{}
\label{init}
\begin{tabular*}{\columnwidth} {ll}
\hline
PLANET & \\
\,\,\,Planet/Star mass & $^{*}\mu=2\times 10^{-3}$\\
\,\,\,eccentricity& $^{*}e_p=0$\\
\,\,\,semi-major axis&$a_p=75\,$AU\\
PARENT BODY RUN &  \\
\,\,\,Number of test particles & $ N_{PB}=2 \times 10^{5}$\\
\,\,\,Initial radial extent & $30<r<130$\,AU (embedded planet case)\\
\,\,\,& $45<r<75$\,AU (inner-disc/outer-planet case)\\
\,\,\,Initial eccentricity & $0\leq e \leq 0.01$\\
\,\,\,Initial surface density & $\Sigma \propto r^{-1}$\\
COLLISIONAL RUNS & \\
\,\,\,Average optical depth & $^{*} \tau= 2\times10^{-3}\,$\\
\,\,\,Number of test particles & $ N_{num}=2 \times 10^{5}$\\
\,\,\,Size range\footnote{$s_{cut}$ is the radiation pressure blow-out size} & $s_{cut}\leq s \leq 40 s_{cut}$\\
\,\,\,Size distribution at $t=0$ \footnote{when released from the parent body population} & $dN(s) \propto s^{-3.5}ds$\\
\hline
\end{tabular*}
\end{minipage}
\end{table}

\section{Results} \label{resu}

\subsection{Pure N-body runs: parent body population} \label{parent}

\begin{figure*}
\makebox[\textwidth]{
\includegraphics[scale=0.39]{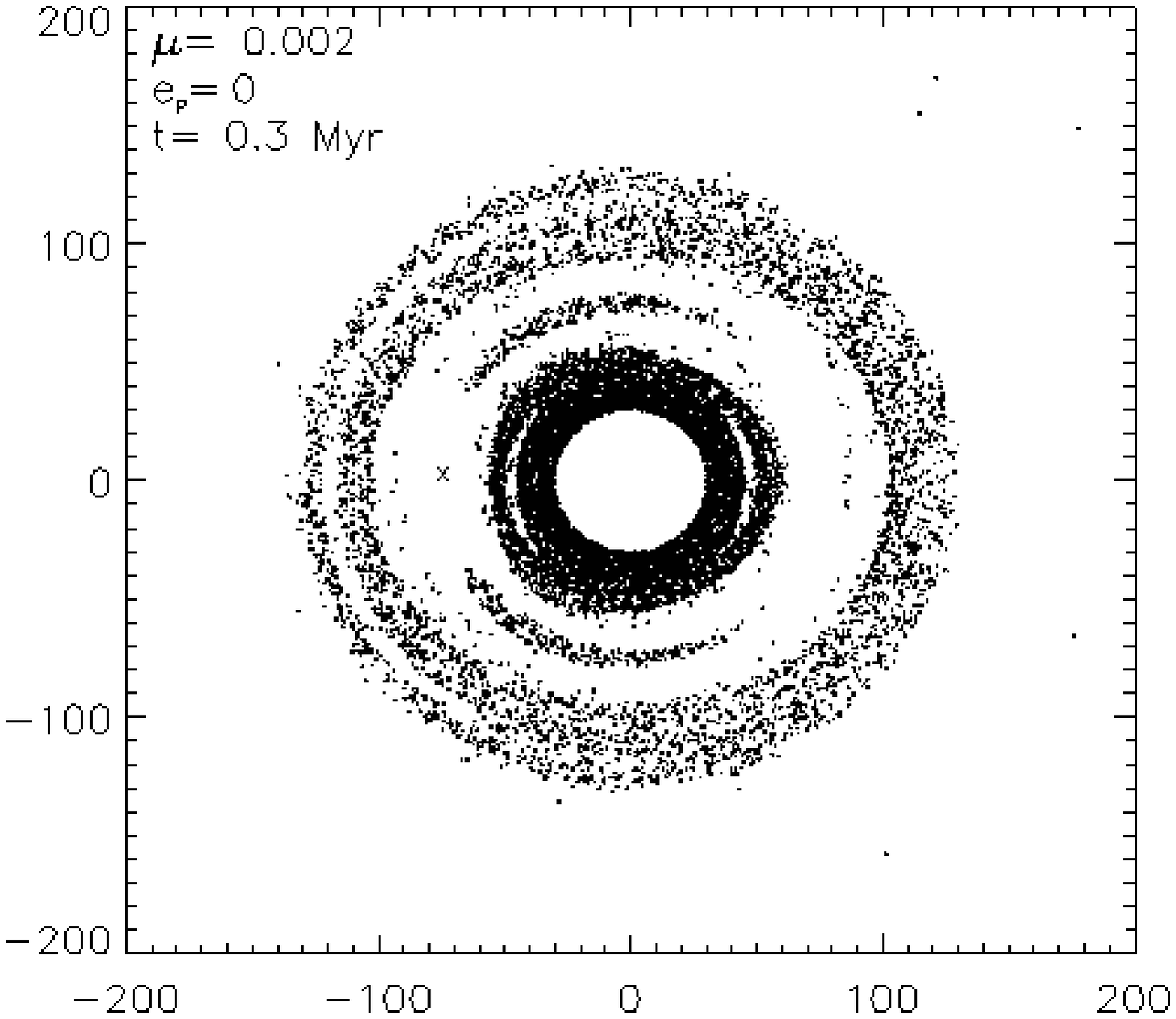}
\includegraphics[scale=0.39]{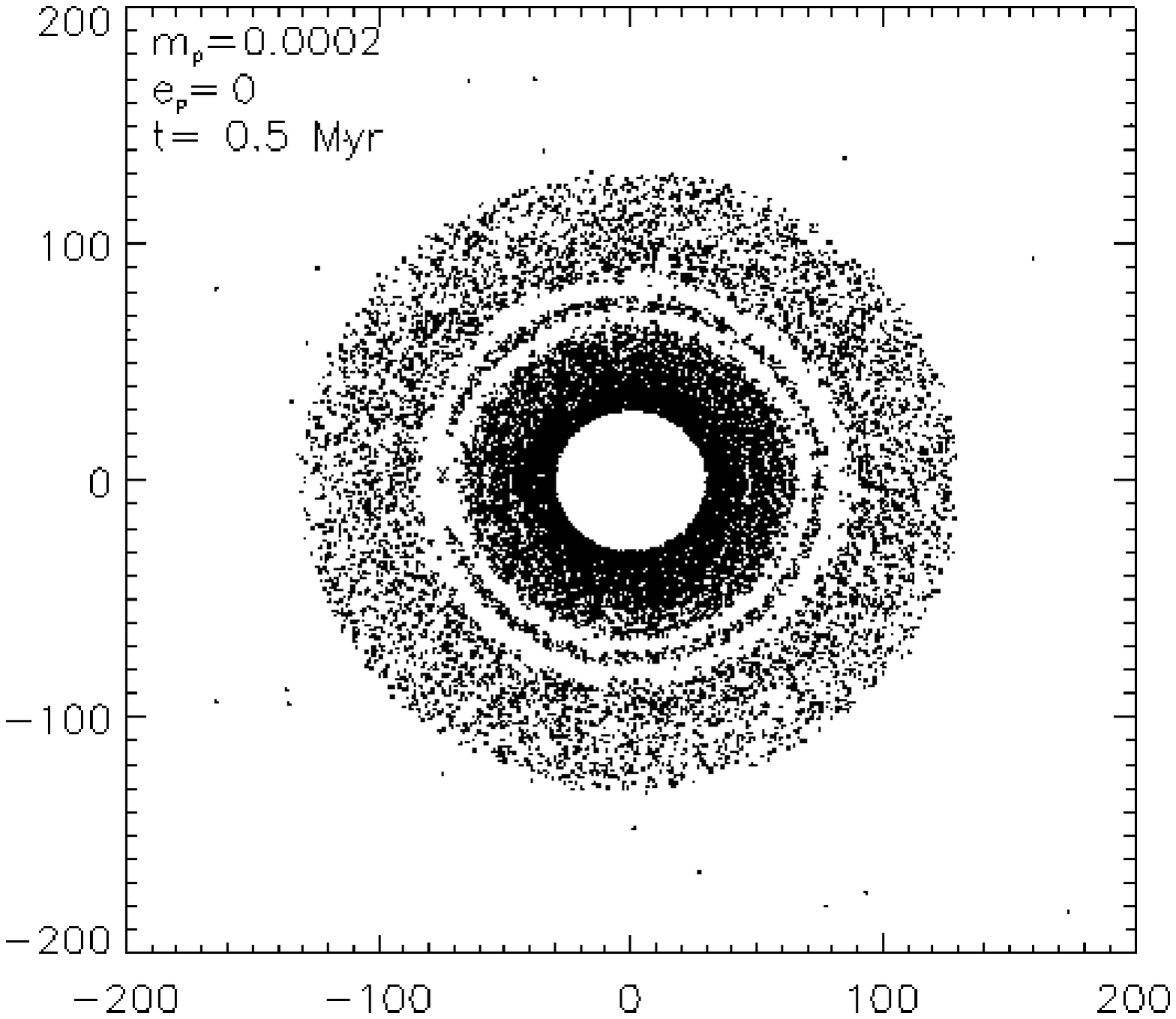}
}
\makebox[\textwidth]{
\includegraphics[scale=0.39]{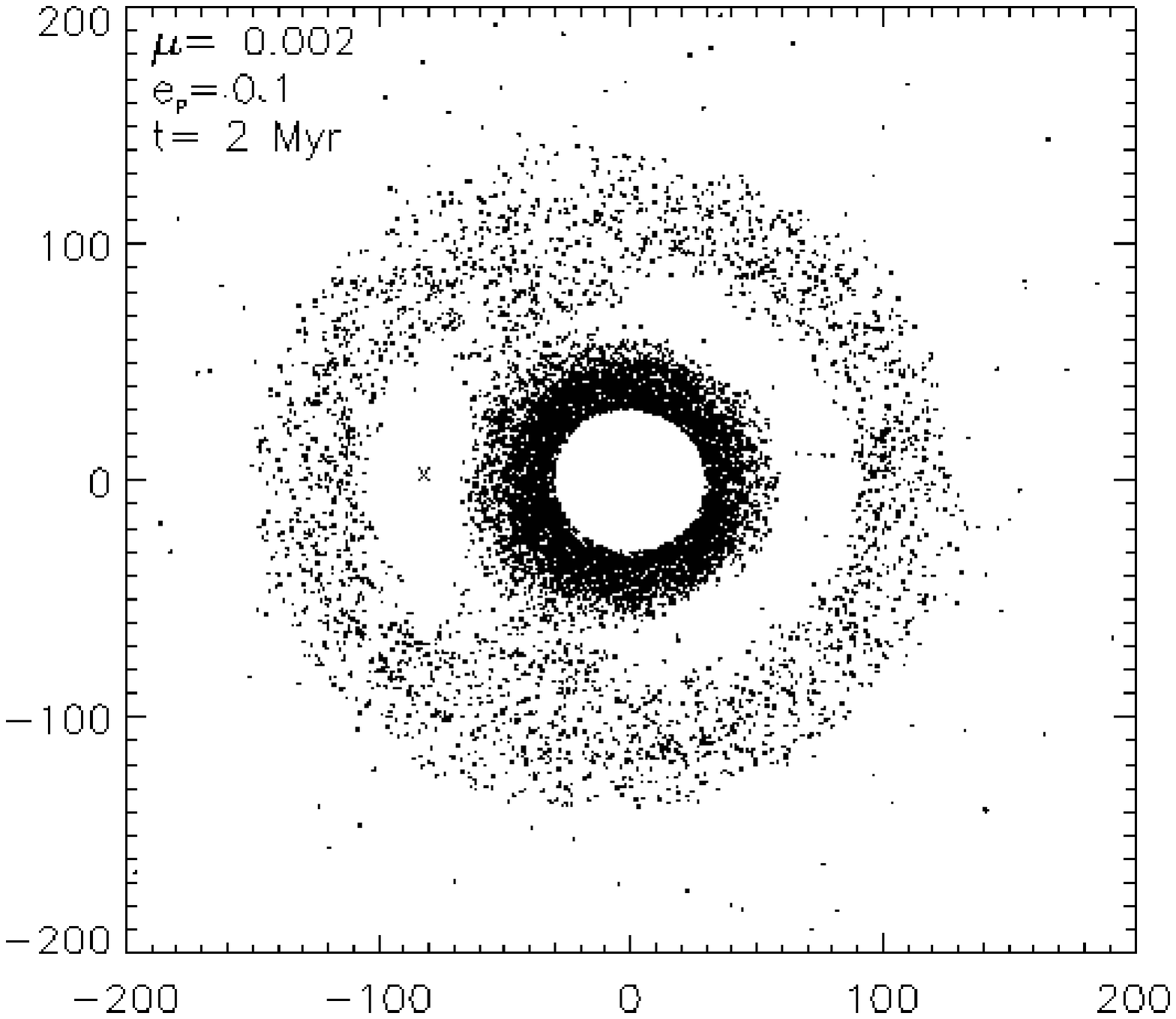}
\includegraphics[scale=0.39]{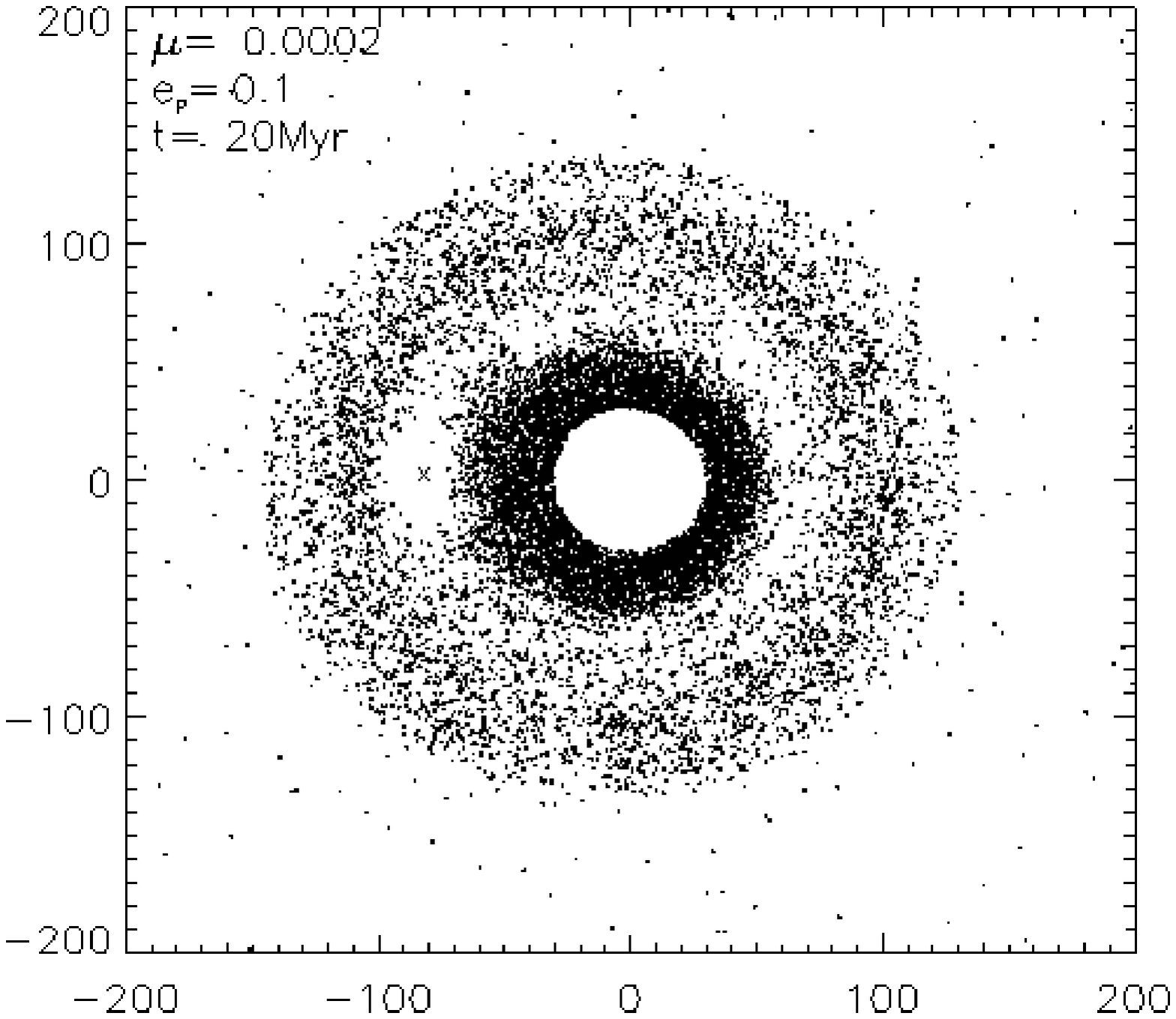}
}
\makebox[\textwidth]{
\includegraphics[scale=0.39]{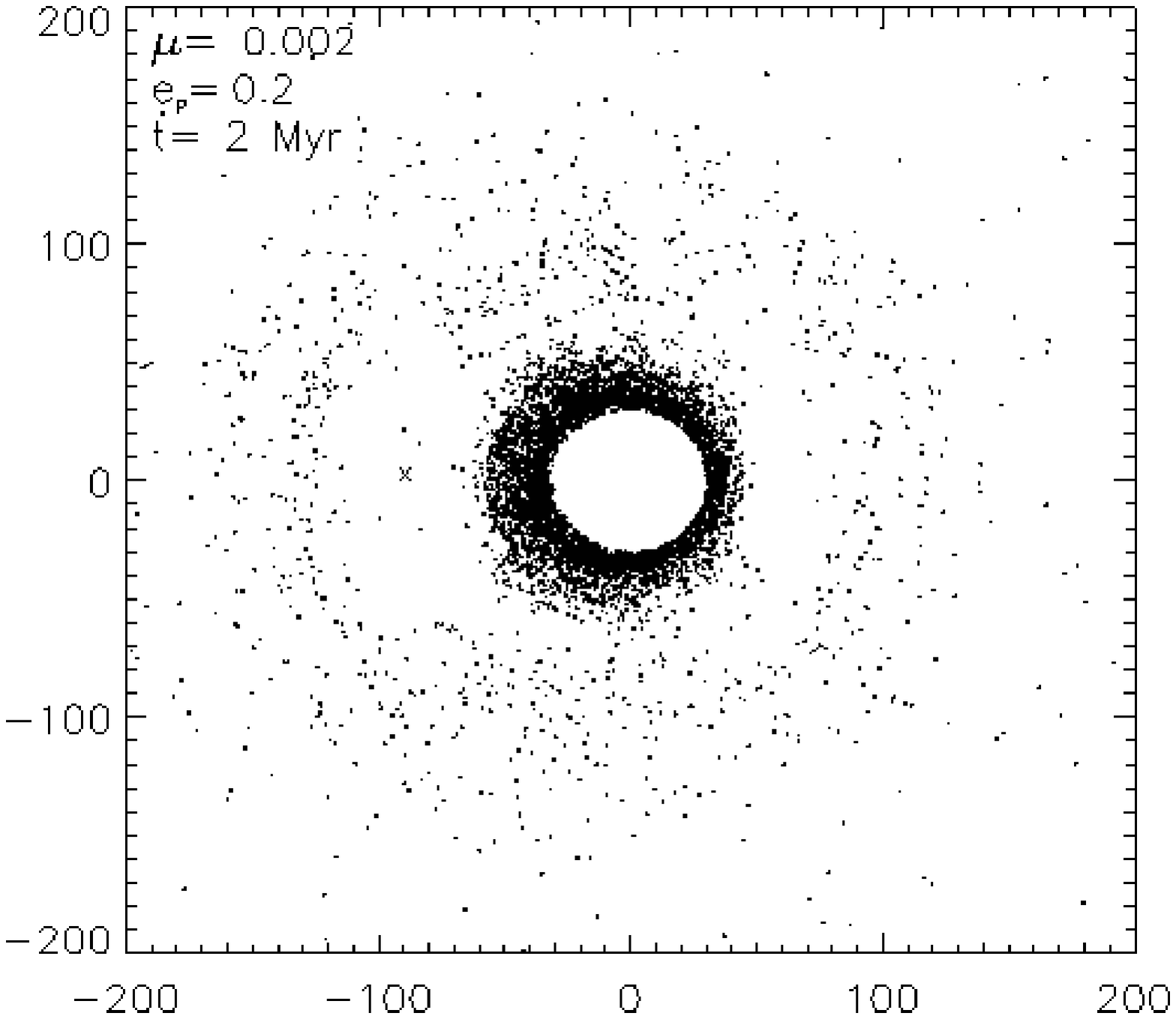}
\includegraphics[scale=0.39]{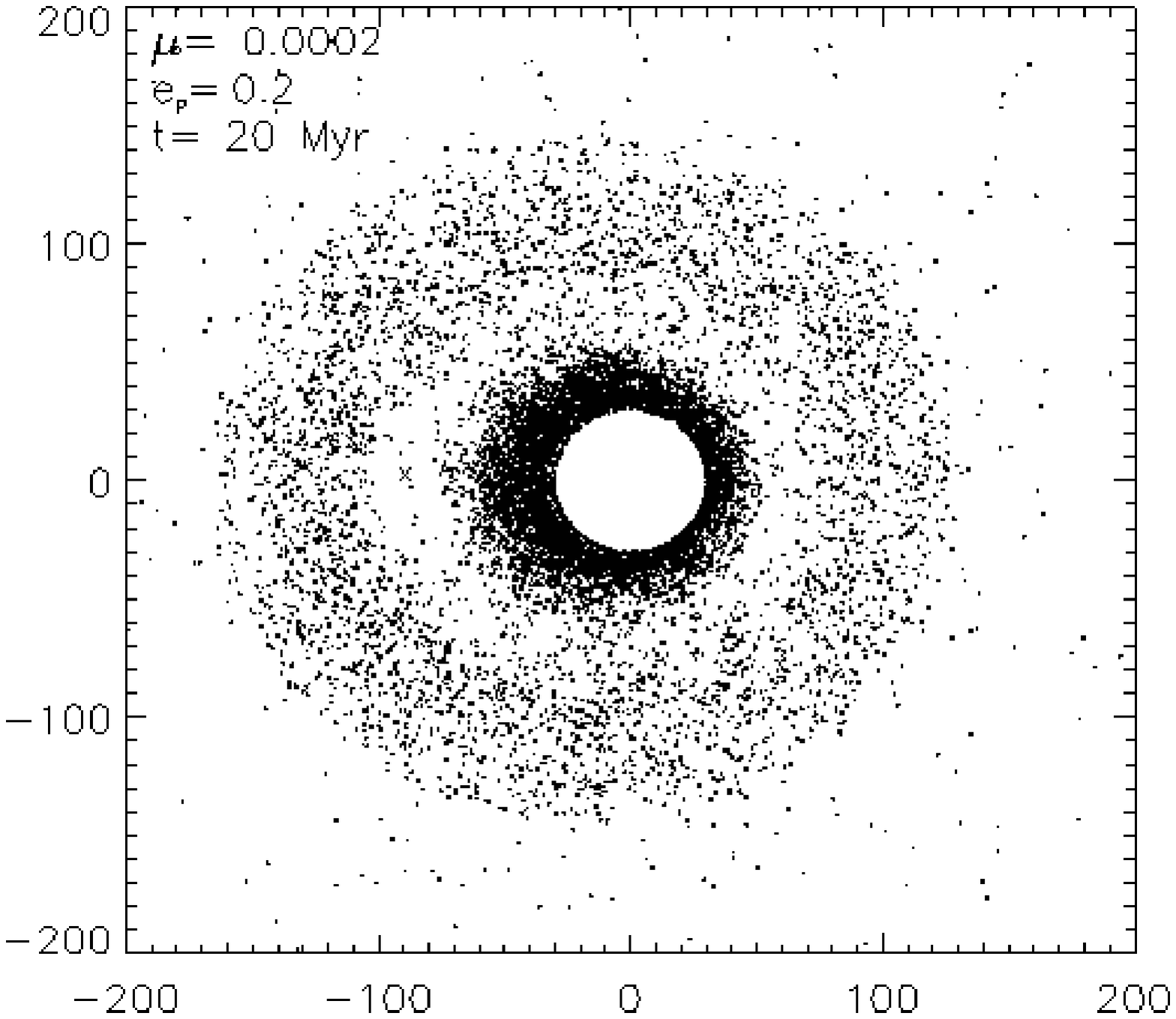}
}
\makebox[\textwidth]{
\includegraphics[scale=0.39]{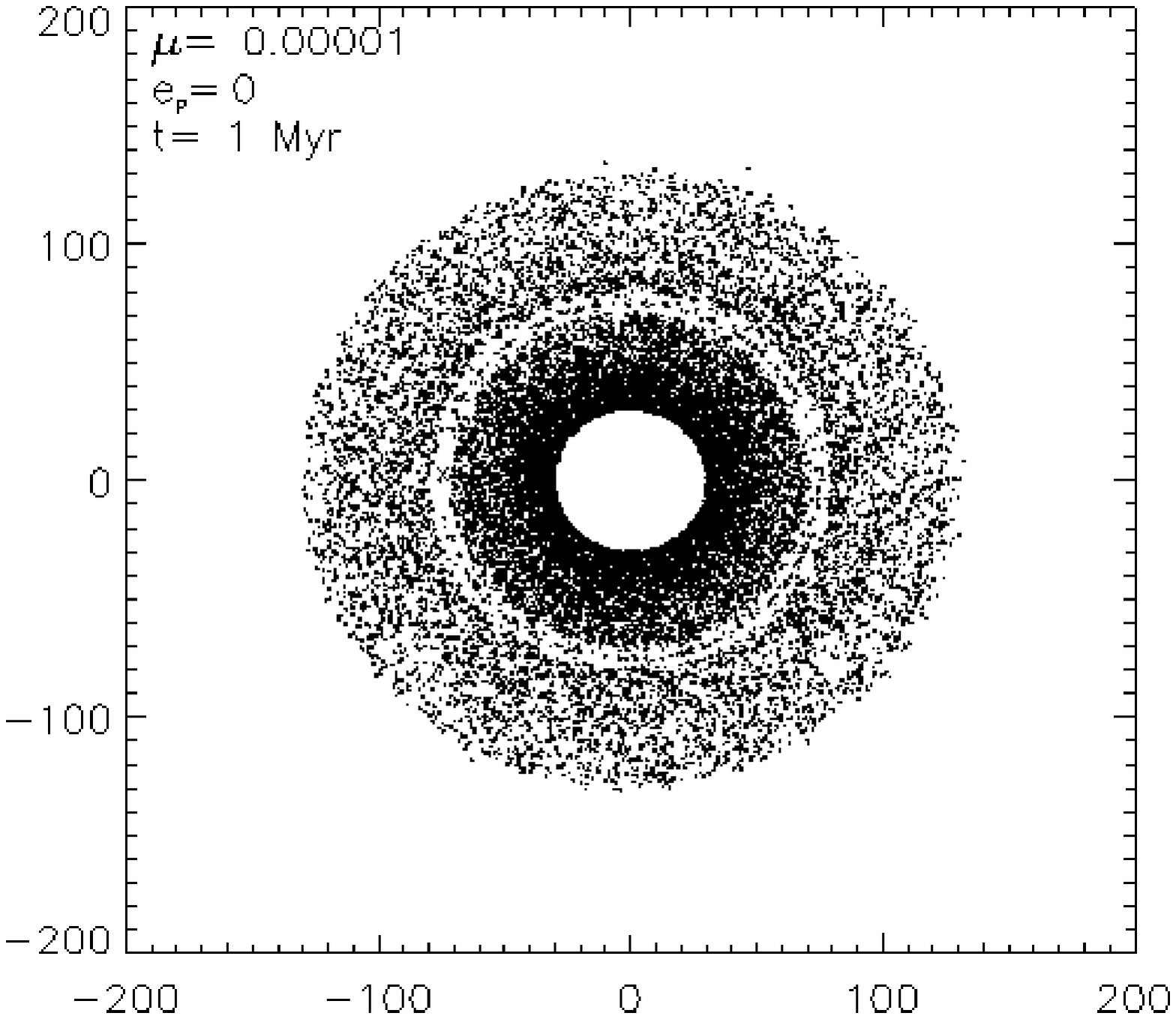}
\includegraphics[scale=0.39]{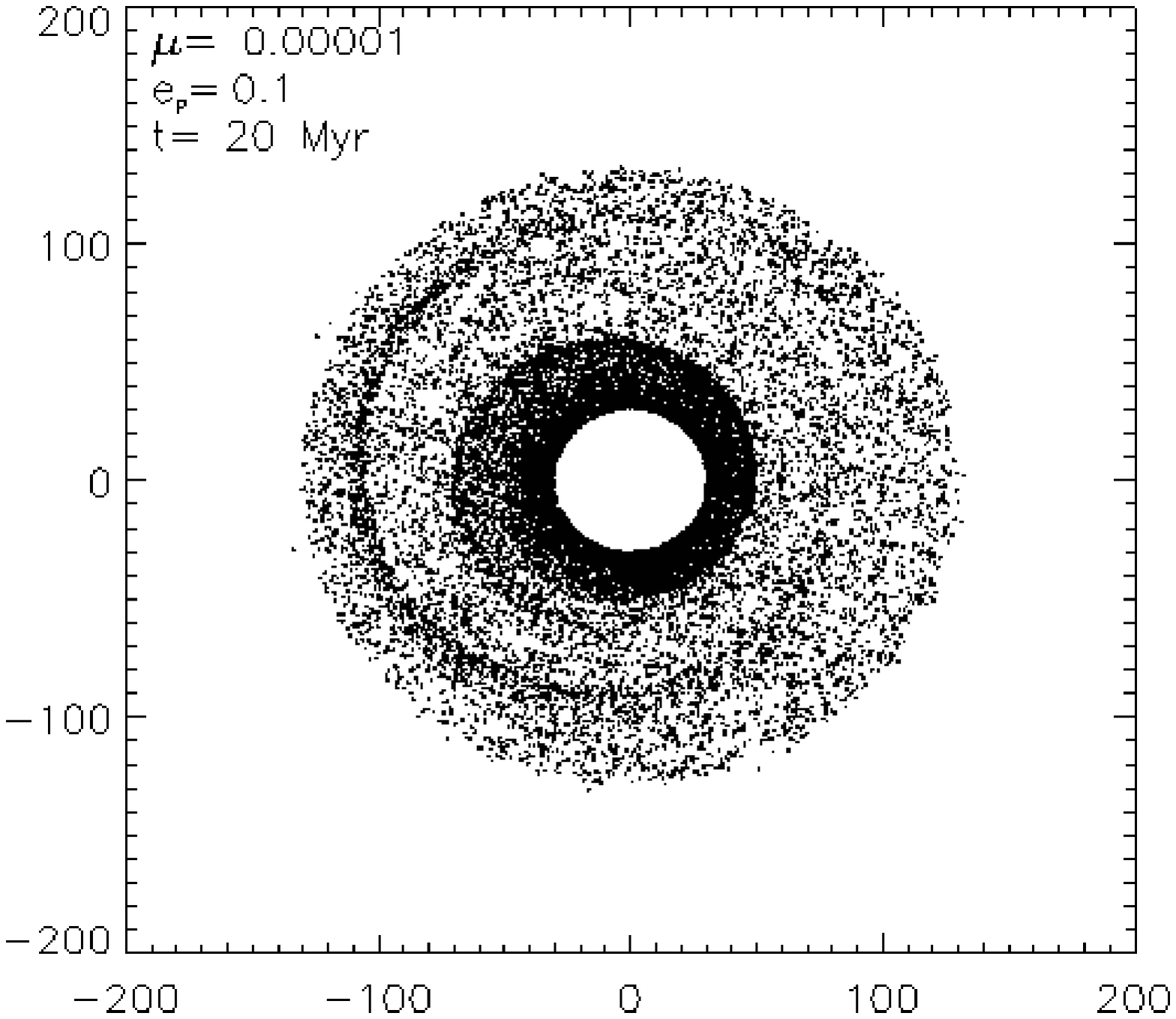}
}
\caption[]{Spatial distribution at the end of the parent body dynamics-only runs, i.e., once a steady state has been reached (except for the $\mu= 10^{-5}$, $e_b=0.1$ run), for different planet orbits and masses. Distances are given in AU. The planet's semi-major axis is 75\,AU for all cases, and is aligned with the X-axis, with its perihelion on the right hand side. All snapshots are shown when the planet passes at aphelion (the planet's location is marked by a cross). 
For all runs except the $\mu=10^{-5}$, $e_b=0.1$ one, the time given is the time at which steady-state has been reached. For the $\mu=10^{-5}$, $e_b=0.1$ run, the time is the one at which simulation has been stopped, i.e., before steady state has been reached (see text for details).
}
\label{PBR}
\end{figure*}

In these purely dynamical runs we consider the evolution of collisionless large particles, not affected by radiation pressure, which will serve as seed particles for the collisional runs. 
As already mentioned, these collisionless simulations are only the first step of our complex numerical procedure and not the purpose of our study, but we will present some of them (for the embedded planet case only) for pedagogical purposes and also because they will serve as easy references to identify the effect of collisions and radiation pressure on the purely dynamical behaviour of perturbed discs. 

For these parent body runs (hereafter PBR), we let the system evolve until a dynamical steady state has been reached,i.e., until the disc's spatial structure does not vary between two passages of the perturbing body at the same orbital phase (see TBO12).  For the case of a planet on a circular orbit, this happens relatively quickly, typically less than a hundred planet orbits, for all the planet masses considered (see Fig.\ref{PBR}a, b and g). For the nominal $\mu=2\times 10^{-3}$ and $e_p=0$ case, several pronounced structures are visible. The most prominent one is the gap in the middle of the disc, corresponding to the chaotic unstable zone surrounding the planet's location \citep[e.g.][]{wisd80}, where only the two stable co-orbital Lagrange points L4 and L5 are populated. This gap corresponds to the so-called "feeding zone" surrounding the planet's orbit.
Clearly visible are also the 1:2 and 2:1 mean motion resonances (hereafter MMR) at 47 and 119 AU, respectively. The apparent gaps at the resonances' location is due to the fact that resonant particles reach high eccentricities $e$ and large radial excursions, thus spending a short fraction of their orbit at the exact radial resonant location.
For a lower planet mass $\mu=2\times 10^{-4}$, the gap logically becomes more narrow, and the whole co-orbital azimutal region is now populated, except for the region just surrounding the planet. 

For an eccencric planet, the system's evolution is very different, and the time to reach steady state can be very long, especially for small perturbers. Indeed, as has been thoroughly investigated by \citet{wyat05}, the dynamical response of a disc to the perturbing presence of a planet with $e_p>0$ is the formation of two spiral waves that propagate outwards and inwards from the planet's position. As time passes, the spirals become more and more tightly wound so that they eventually become unnoticeable and appear as an asymmetric ring. The time at which this disapearance occurs depends on radial distance to the planet and is roughly comparable to the secular precession time \citep{wyat05}
\begin{equation}
t_{sec} = 6.15 \left(\frac{a_{p}}{a}\right)^{-2.5} \alpha_{p}^{2}  \left(\frac{1}{b_{3/2}^{1}(a_{p}/a)}\right)\,\,t_{sec(3:2)},
\end{equation}
where $\alpha_{p}=1$ for $a_{p}<a$ and $\alpha_{p}=a_{p}/a$ for $a_{p}>a$, $b_{3/2}^{1}(a_{p}/a)$ is the Laplace coefficient (see \citet{wyat05} for more details), and $t_{sec(3:2)}$ is the secular timescale at the location of the 3:2 mean motion resonance:
\begin{equation}
t_{sec(3:2)} = 0.651 t_{p} \left(\frac{M_{p}}{M_\star}\right),
\end{equation}
where $t_{p}$ is the orbital period of the planet.

In agreement with \citet{wyat05}'s results, we witness the development and progressive disappearance of these 2 spirals. The dynamical steady state of the system is reached when secular precession has rendered the spiral structures indiscernable everywhere in the system, i.e., when $t>$max($t_{sec}(a_{in}),t_{sec}(a_{out})$). For our reference case of a $\mu = 2\times 10^{-3}$ planet, and with $a_{in}=30\,$AU and $a_{out}=130\,$AU this happens after $\sim 2\times 10^{6}$years. For a smaller $\mu = 2\times 10^{-4}$ planet, this time to steady-state increases to $\sim 2\times 10^{7}$years (Fig.\ref{PBR}d and f). It is important to point that this value exceeds the age of some bright and young debris discs such as HR4796 or $\beta$ Pictoris. We adopt a conservative approach and exclude from our study all cases for which steady state is reached on timescales longer than this value. We show however, for illustrative purposes, one such case, with $e_p=0.1$ and $\mu =10^{-5}$ (Fig.\ref{PBR}h), where the transient spirals are still clearly visible at $\sim 2\times 10^{7}\,$years.

Regarding the spatial distribution of the steady state disc for these eccentric planet cases, we see that, even for a massive $\mu=2\times 10^{-3}$ planet, almost no resonant structure is longer visible in the inner or outer discs, and the Lagrange points, although still present, are much less pronounced, especially in the $e_p=0.2$ case. In addition, the outer disc gets strongly depleted while the inner disc assumes an asymmetric shape aligned with the planet's orbit.

\subsection{Collisional system: embedded planet} \label{sscoll}

\begin{figure*}

\makebox[\textwidth]{
\includegraphics[scale=0.4]{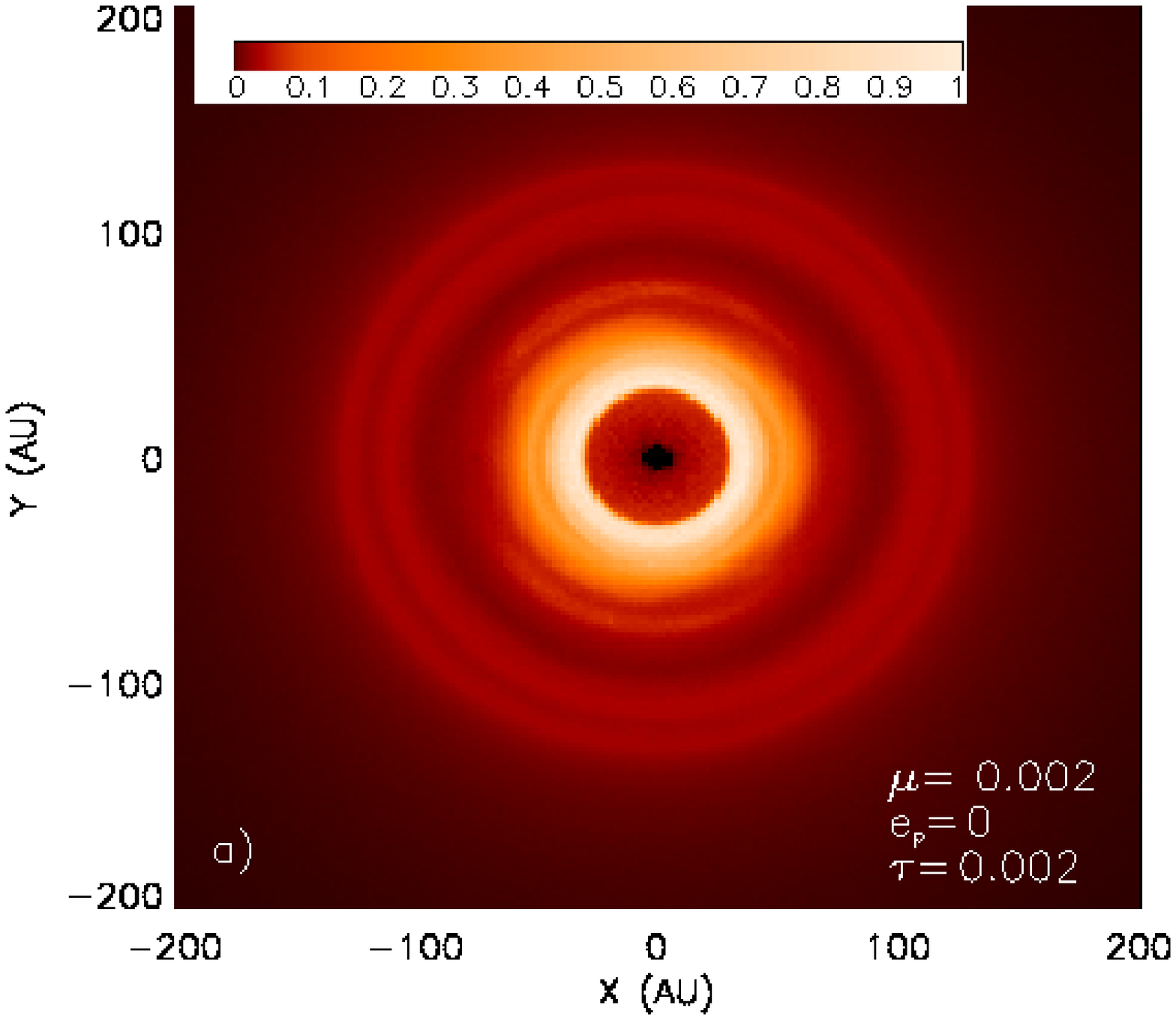}
\includegraphics[scale=0.4]{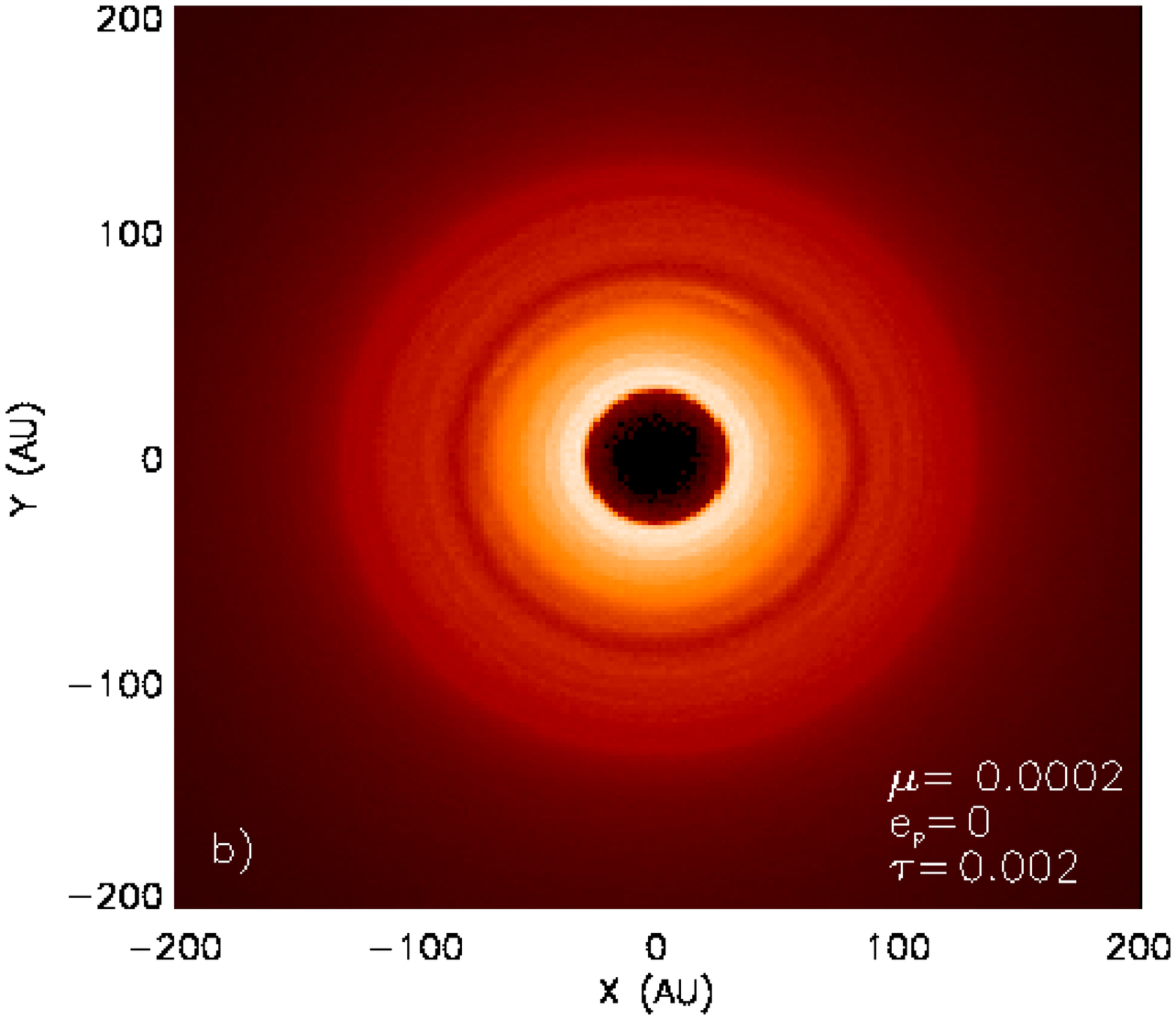}
}
\makebox[\textwidth]{
\includegraphics[scale=0.4]{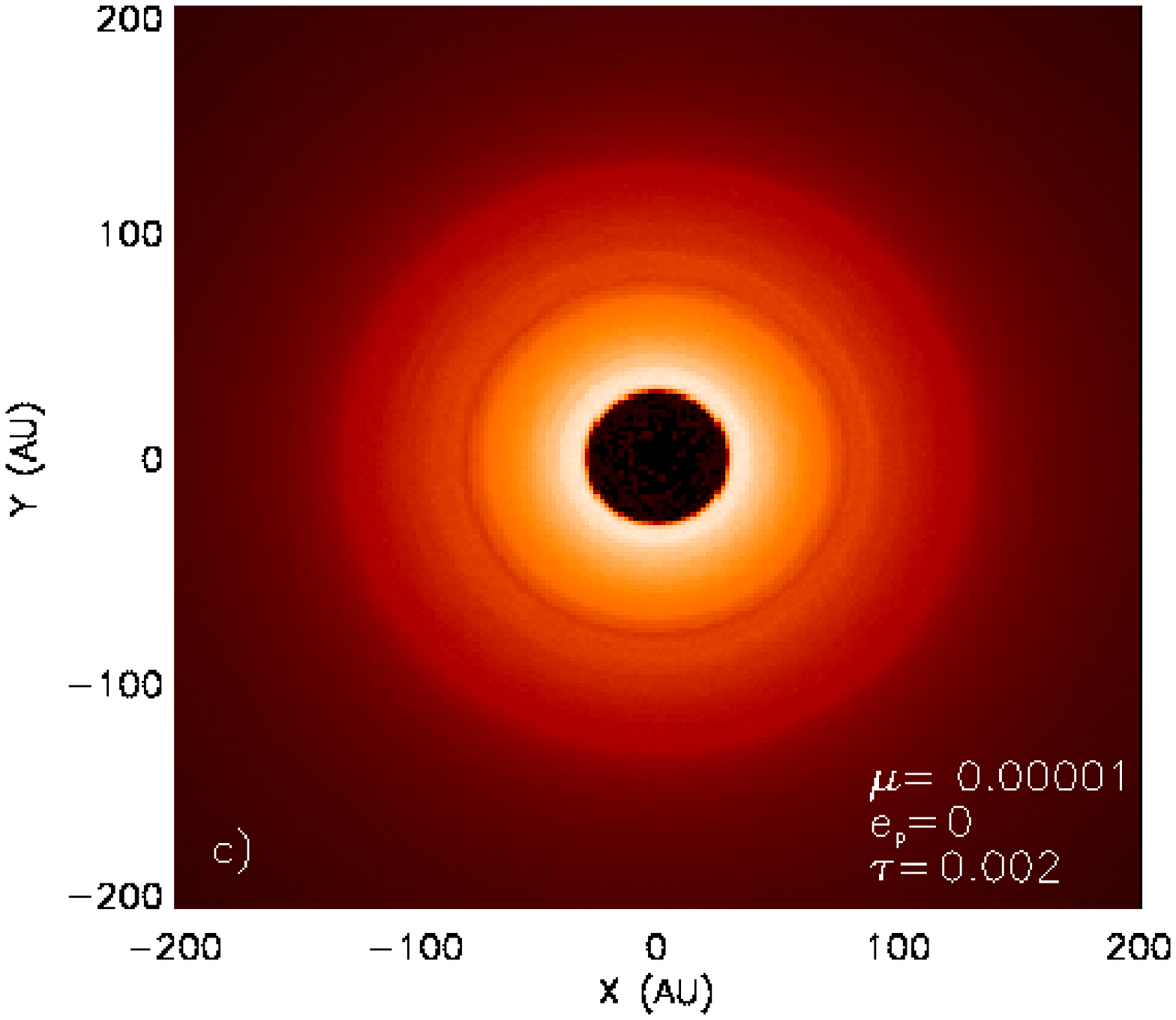}
\includegraphics[scale=0.4]{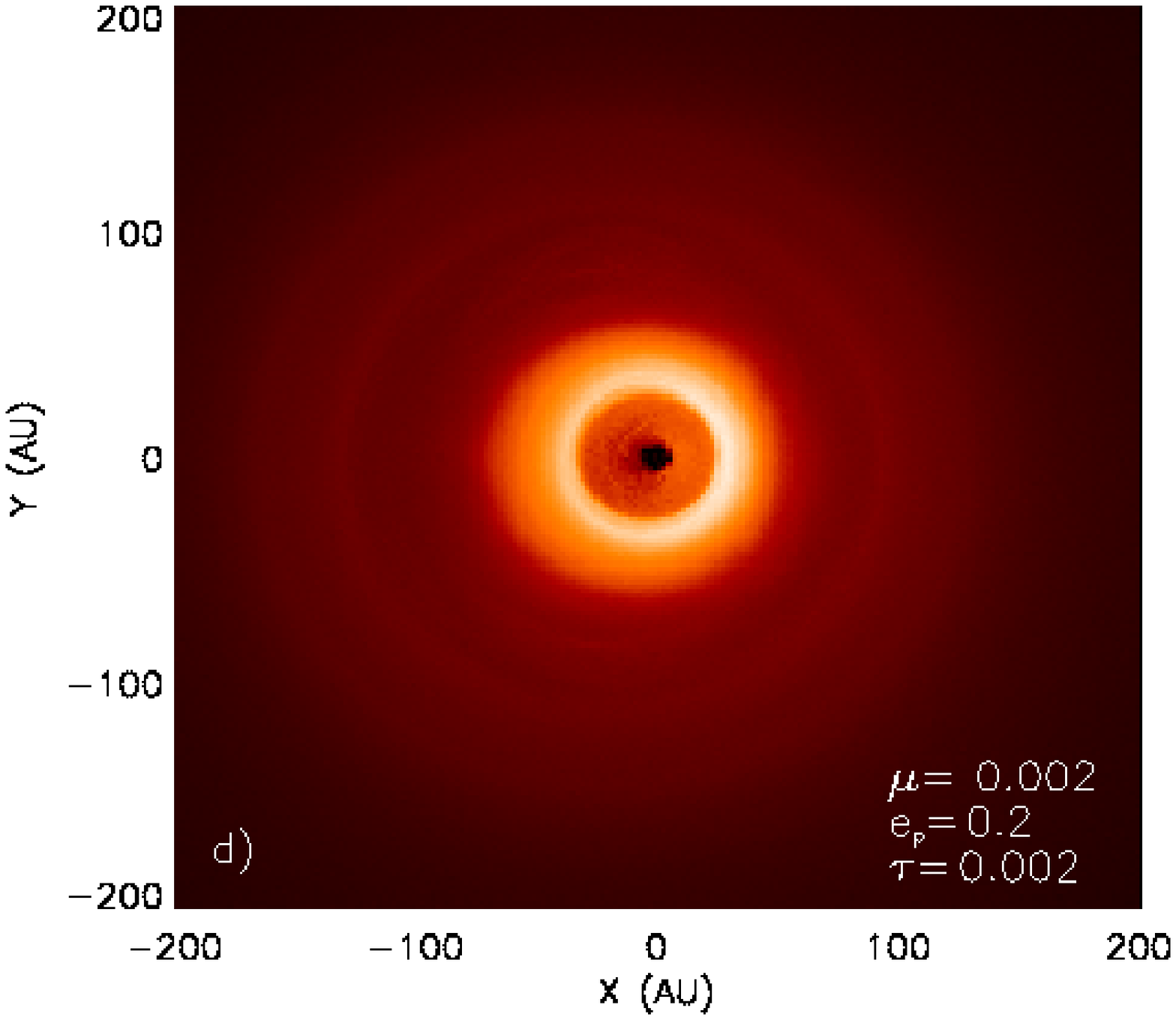}
}
\makebox[\textwidth]{
\includegraphics[scale=0.4]{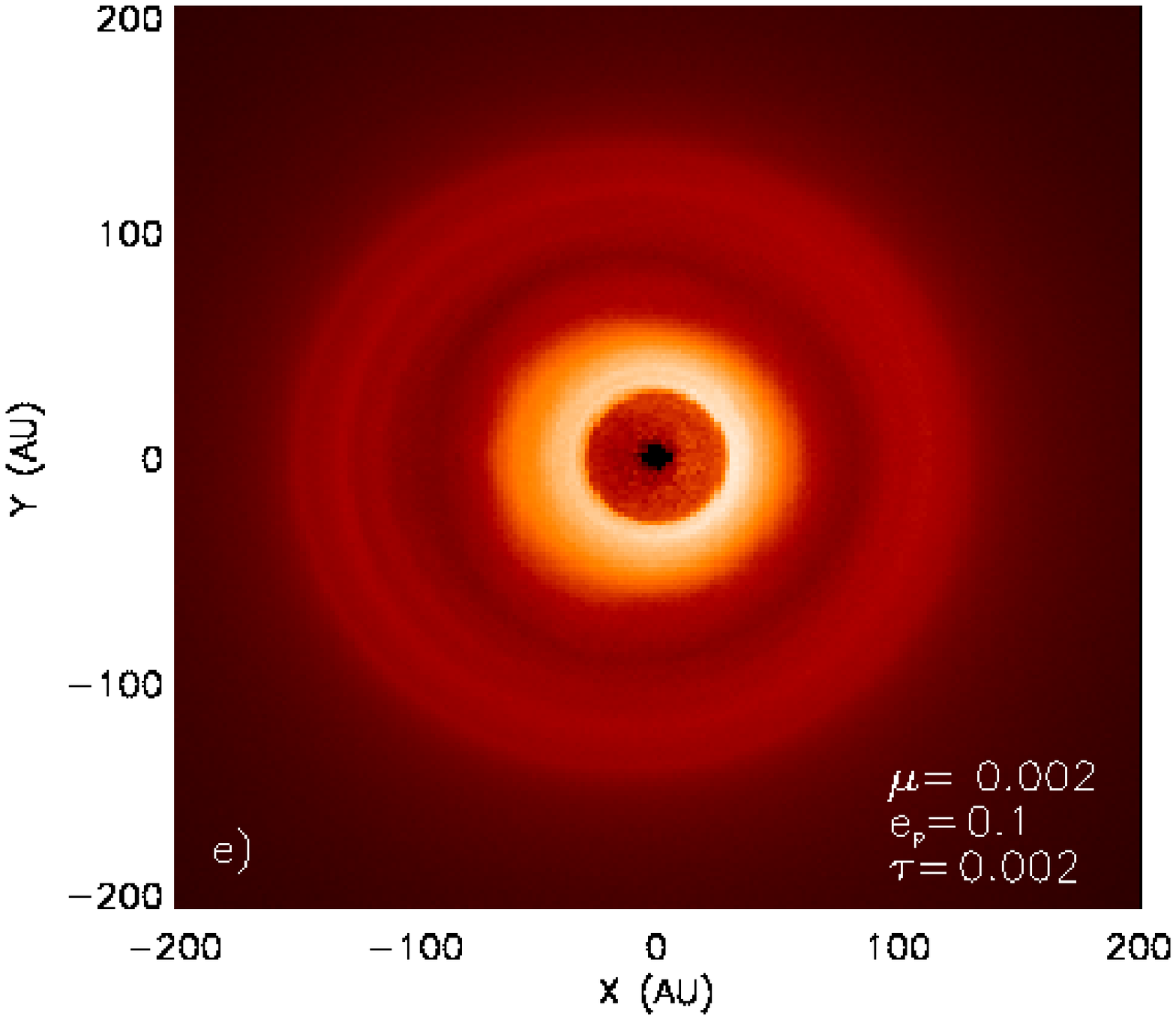}
\includegraphics[scale=0.4]{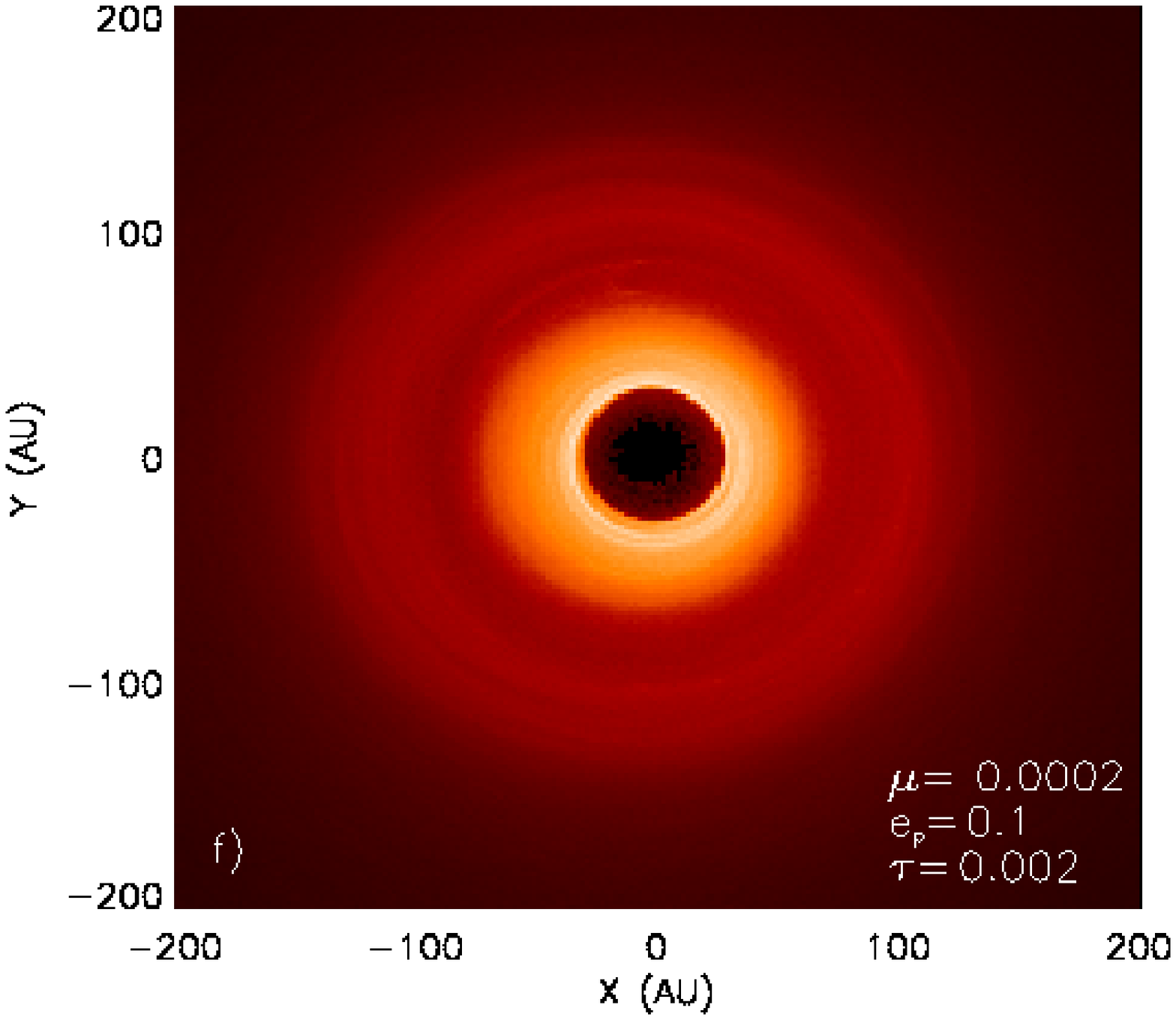}
}
\makebox[\textwidth]{
\includegraphics[scale=0.4]{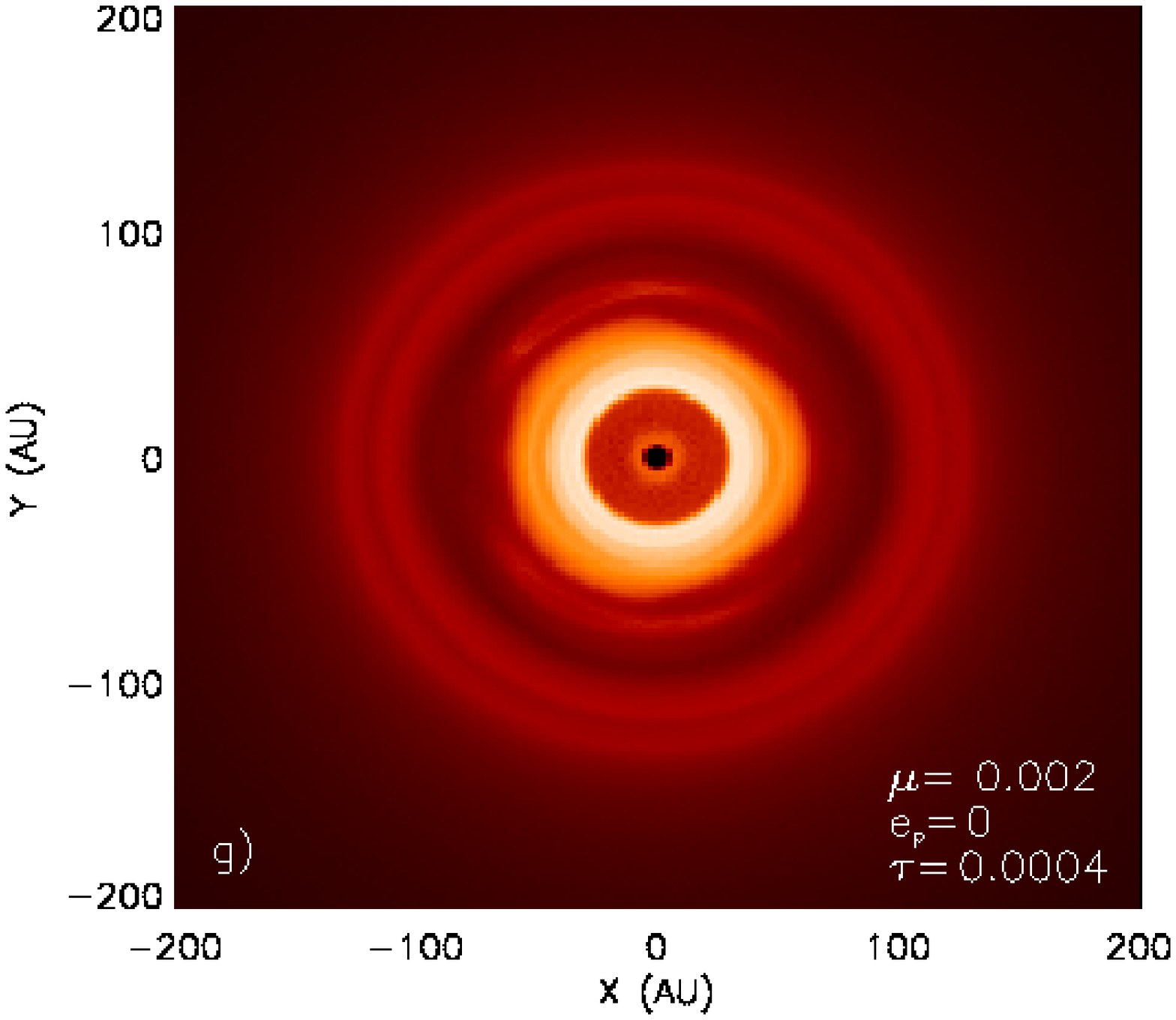}
\includegraphics[scale=0.4]{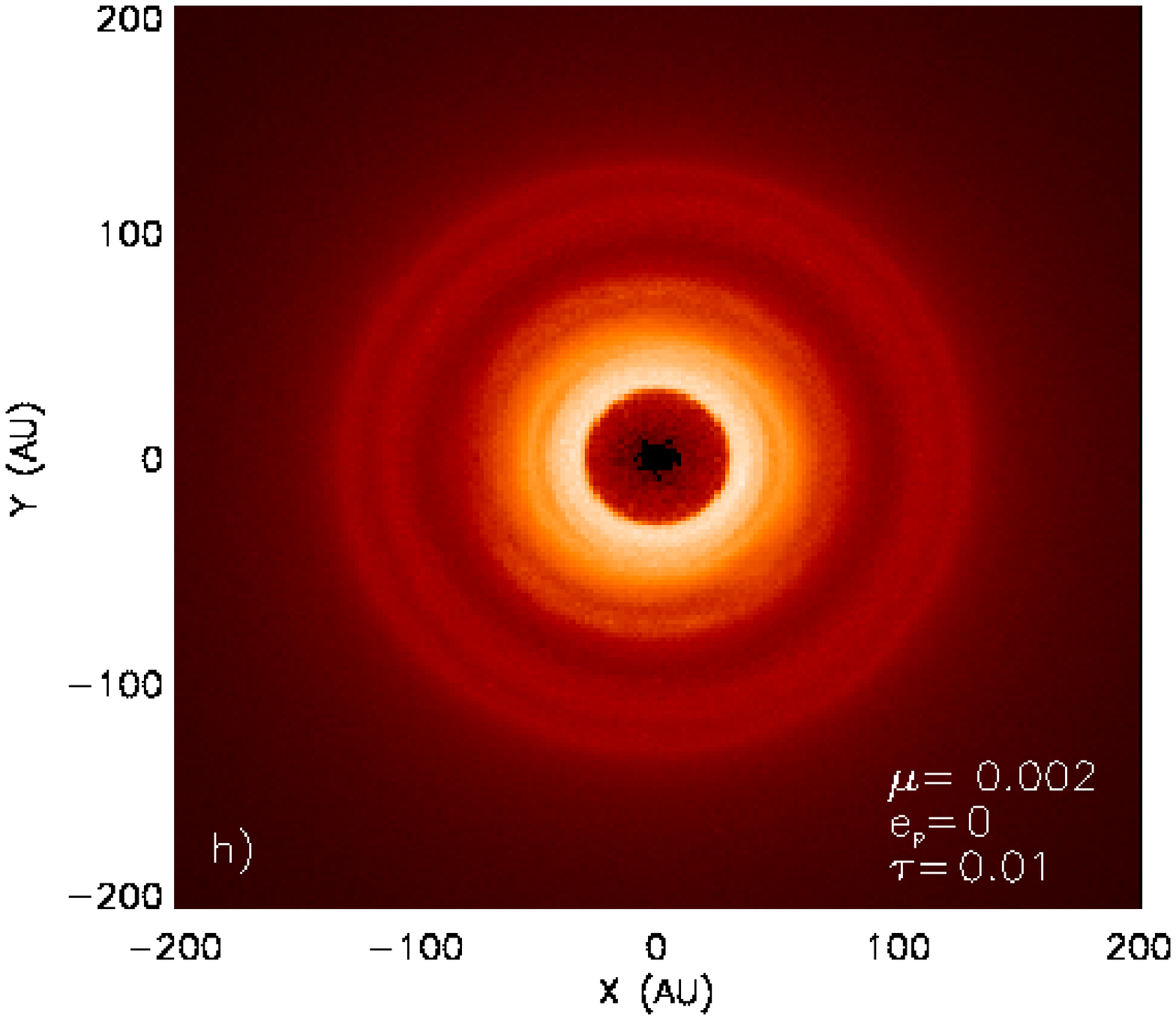}
}
\caption[]{Normalized synthetic images in scattered light, at collisional and dynamical steady state, for different planet and disc configurations (see text for details).
}
\label{collrun}
\end{figure*}

Fig.\ref{collrun} shows synthetic head-on images, in scattered light, at collisional and dynamical steady-state, for different planet orbits and collisional activities in the disc (as parameterized by its average optical depth $\tau_0$). 
For our nominal $\tau_0=2\times 10^{-3}$ case, the main structures observed in the pure N-body runs, such as for example the Lagrangian points and the 2:1 resonances in the $e_p=0$ run, are still clearly visible, but they are much less pronounced. The other consequence of the coupled effect of collisions and radiation pressure is that it injects small, high-$\beta$ grains in regions that are devoid of parent bodies. These regions are the external region beyond the outer edge (at 130\,AU) of the parent body disc, and, more importantly, the dynamically unstable region (the gap) close to the planet's orbit. 
The global effect of collisions and radiation pressure is thus to tend to homogenize the surface density profile, as is clearly illustrated for runs with higher optical depth (i.e., collisional activity), where the differences with the collisionless N-body simulation get much more pronounced. For the $e_p=0$ orbit, the accumulation around the Lagrangian points is for example replaced by a horseshoe co-orbital ring for higher value of $\tau_0$ (Fig.\ref{collrun}h). Conversely, for lower values of $\tau_0$, the spatial profile becomes much closer to that of the parent bodies (Fig.\ref{collrun}g).

Despite this blurring effect of collisions, for all the head-on images displayed in Fig.\ref{collrun}, the planet's radial location remains relatively easy to determine from the position of the radial gap. For high-mass low $e_p$ cases, even the planet's azimutal position remains easy to spot in between the well defined Lagrangian points, with the noticeable exception of the high $\tau_0$ case (Fig.\ref{collrun}h) where the steady production of small collision fragments erases almost all azimutal structures in the co-orbital region.

\begin{figure}
\includegraphics[width=\columnwidth]{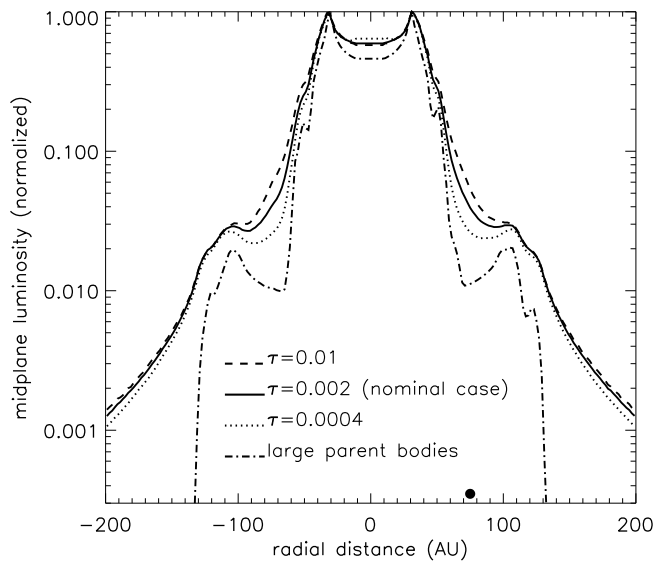}
\caption[]{Synthetic midplane luminosity profiles for a disc seen edge-on, in scattered light, for the standard $\mu=2\times 10^{-3}$ and $e_p=0$ case and for different values of the disc's average optical depth (i.e., collisional activity). The shown profiles correspond to a cut along the X axis when the planet passes at its maximum elongation along this axis (the planet's location is marked by a black circle). Profiles are computed assuming a grey scattering function. The profile for the large parent bodies is directly derived from the PB runs presented in Section \ref{parent}.
}
\label{cedgec}
\end{figure}

The situation becomes very different if the system is seen \emph{edge-on}. Fig.\ref{cedgec} shows the midplane luminosity profile along a radial cut passing by the planet's current position. While the distribution of large parent bodies shows a pronounced gap around the planet's radial position (Fig.\ref{cedgenc}), with the flux dropping by a factor $\sim 5$ in the 50-90AU region, the coupled effect of collisions and radiation pressure on small grains greatly alters this picture. Even for a relatively low $\tau_0 = 4\times 10^{-4}$ value, the luminosity drop around the planet falls to less than a factor 2. It is of less than 30\% in the nominal $\tau_0 = 2\times 10^{-3}$ case, and falls down to a few percent for even more collisionally active discs ($\tau_0=0.01$). Furthermore, this small flux variation is not peaked at the planet's location but diluted over a rather extended region, so that the planet's effect on the profile is a wide and poorly defined knee. 

\begin{figure}
\includegraphics[width=\columnwidth]{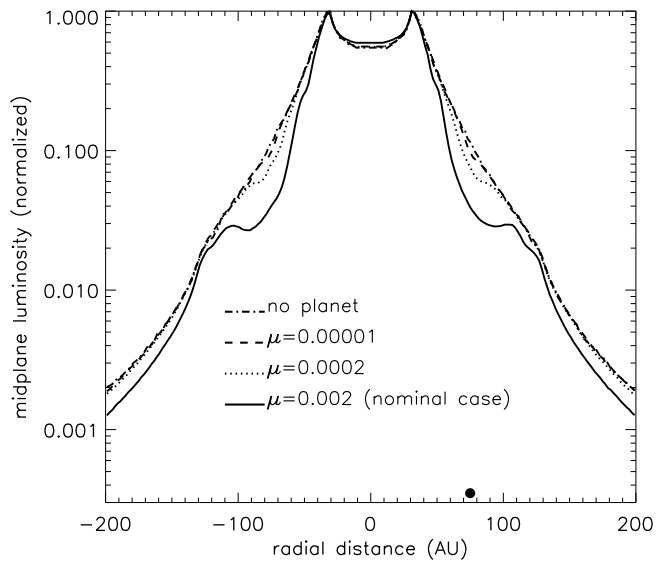}
\caption[]{Same as Fig.\ref{cedgec}, but with $e_p=0$ and $\tau_0=2\times 10^{-3}$ and when varying the mass of the perturbing planet.
}
\label{cedgem}
\end{figure}

\begin{figure}
\includegraphics[width=\columnwidth]{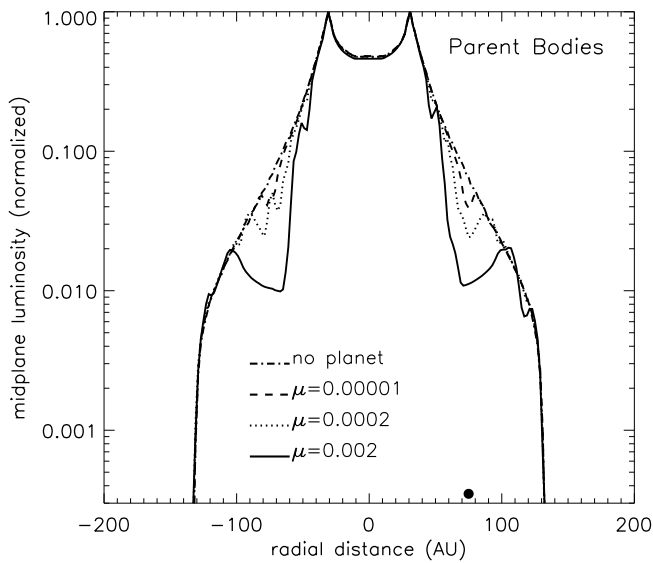}
\caption[]{Same as Fig.\ref{cedgem}, but for the large parent bodies ($e_p=0$).
}
\label{cedgenc}
\end{figure}

The effect of the planet on the luminosity profile is even weaker when considering masses smaller than the $\mu=2\times 10^{-3}$ super-Jupiter of our reference case. As can be clearly seen in Fig.\ref{cedgem}, for a  dense debris disc ($\tau_0=2\times 10^{-3}$), a Saturn-like planet ($\mu=2\times 10^{-4}$) leaves a barely perceptible signature, and a super-Earth ($\mu=10^{-5}$) results in a radial profile almost identical to the no-planet case. This is in stark contrast to the profile that would have been obtained had collisions and radiation pressure been neglected. Fig.\ref{cedgenc} shows such profiles, derived from the parent body runs. As can be seen, in these fiducial cases, there is always a well defined and narrow gap close to the planet's radial location, even for the smallest case of a $\mu= 10^{-5}$ super-Earth. This clearly illustrates how the coupling of collisional activity and radiation pressure efficiently erases the signature of a planet (see Sec.\ref{Discu} for further discussion).

\begin{figure}
\includegraphics[width=\columnwidth]{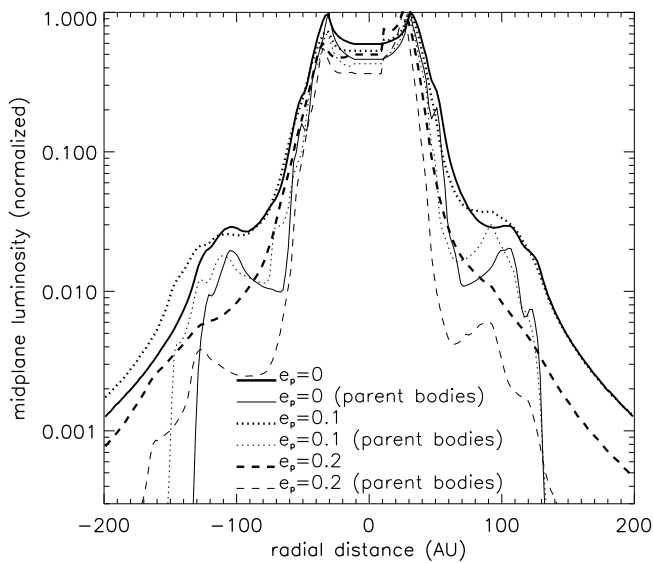}
\caption[]{Same as Fig.\ref{cedgec}, but for different eccentricities of the planet's orbit (its mass being $\mu=2\times 10^{-3}$).
}
\label{cedgeex}
\end{figure}

The blurring effect of collisions is even more apparent for an eccentric embedded planet. For the $e_p=0.1$ case, collisional production of small high-$\beta$ grains homogenizes the region co-orbital with the planet (Fig.\ref{collrun}e), in sharp contrast to the pure N-body runs where the co-orbital region is almost empty, safe for the 2 Lagrangian points. 
The same effect is observed for the $e_p=0.2$ case, where no azimutal inhomogeneity is visible in the co-orbital region on the 2-D map (Fig.\ref{collrun}d). For this specific case, however, the collisional runs confirm the main result of the parent-body runs, i.e., a strong depletion in the region beyond the planet's orbit. This depletion is so strong that it almost completely erases the frontier between the planet's co-orbital region and the outer disc.
If the system is seen edge-on, then the signature of an eccentric planet becomes even weaker (Fig.\ref{cedgeex}). It is even less pronounced than in the $e_p=0$ case. For the $e_p=0.2$ run, there is almost no variation in the radial luminosity profile at the planet's distance. Here again, this result is in sharp contrast to the distribution of large parent bodies, for which, regardless of the value of $e_p$, a clear dip is always visible in the edge-on luminosity profile.

\subsection{Collisional system: Inner disc/external planet} \label{extplan}

\begin{figure}
\includegraphics[width=\columnwidth]{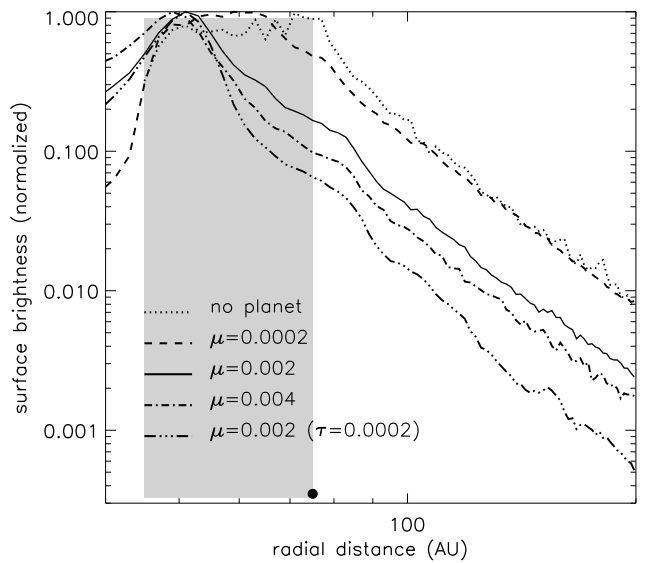}
\caption[]{Inner-ring/outer-planet configuration. Steady-state synthetic surface brightness, when the system is seen \emph{head on}, along a radial cut passing by the planet's current location. The grey area marks the initial radial extent of the parent body ring.
}
\label{cint}
\end{figure}

\begin{figure}
\includegraphics[width=\columnwidth]{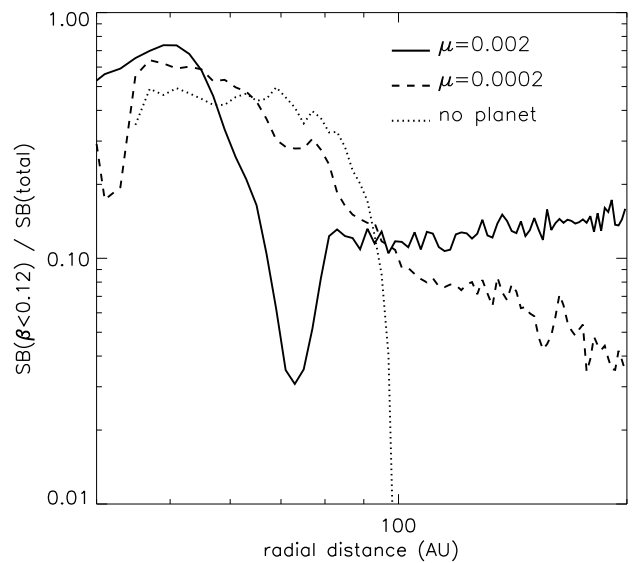}
\caption[]{Inner-ring/outer-planet configuration. Relative contribution of the "big" grains ($\beta\leq0.12$) to the total luminosity for 3 different cases.
}
\label{cintf}
\end{figure}

\begin{figure}
\includegraphics[width=\columnwidth]{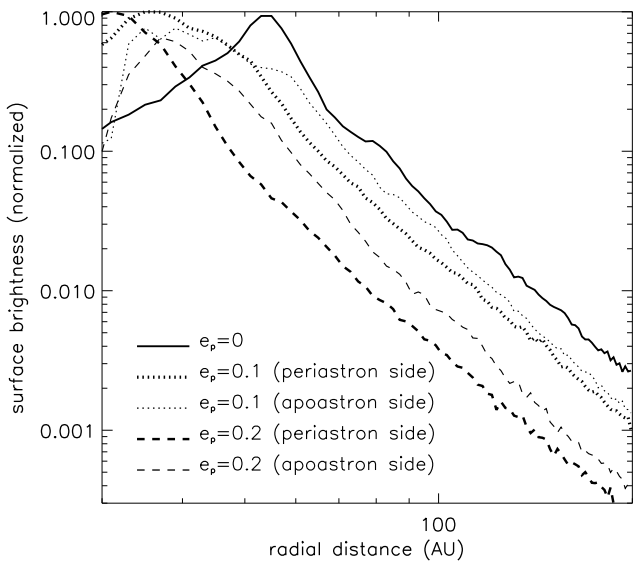}
\caption[]{Same as Fig.\ref{cint}, but for different eccentricities of the exterior planet (and for $\mu=0.002$ and $\tau_0=0.002$). For the $e_p > 0$ cases, plots are presented for a passage of the planet at periastron and apoastron. Note that, since for these $e_p > 0$ cases the unstable chaotic region inside the planet's orbit is much wider than when $e_p=0$, the initial disc of parent bodies extended down to 25\,AU for the $e_p=0.1$ and $e_p=0.2$ runs.
}
\label{cintex}
\end{figure}

We now consider a different configuration, where the planet is located exterior to an inner debris ring. The main objective is here to see to what extent the planet can shape the outer edge of the disc, and in particular if it can affect the luminosity profile in $\propto r^{-3.5}$ that should "naturally" form beyond the radial location of a collisional debris ring \citep{theb08}.
We consider a standard set-up where the inner ring of parent bodies initially extends from 45 to 75\,AU and the planet is positioned at the outer edge of the ring, i.e., 75\,AU. We restrict ourselves to the case of a planet on a circular orbit, and explore its mass and the disc's collisional activity as free parameters.

Fig.\ref{cint} shows a radial cut of the surface brightness profiles when the system is seen head-on. The first effect of the planet is to truncate the disc of parent bodies, at a distance corresponding to its chaotic zone. For each different curve, the location of the outer edge of the stable parent body ring corresponds approximately to the transition point where the brightness peaks before rapidly decreasing with radial distance, i.e., at $r\sim 61\,$AU for a $\mu=2\times 10^{-4}$ planet, $\sim 52\,$AU for a $\mu=2\times 10^{-3}$ planet and $\sim50\,$AU for a $\mu=4\times 10^{-3}$ one.
Interestingly, beyond this outer edge $r_{PB}$ of the PB disc, the brightness profile has a slope that only weakly depends on the planet's mass. For the no-planet run, beyond a transition region extending from 75 to $\sim 95\,$AU, it assumes the standard -3.5 value due to the steady collisional production of radiation-pressure sensitive small grains \citep[see][]{theb08}. For a low-mass planet of $\mu=2\times 10^{-4}$ this profile is basically unchanged, and it only gets  slighly steeper, $\sim -4$ to $-4.2$ for more massive planets ($\mu =2\times 10^{-3}$ and $\mu=4\times 10^{-3}$).

These results might seem counter-intuitive, because the $r>r_{PB}$ domain is by definition dynamically unstable and we would expect the planet to eject a large fraction of the small grains, produced in the main ring, that populated this outer region. However, this ejection process takes time, basically needing a close-encounter with the planet, so that small grains can have a long time of residency in this dynamically "forbidden" region. True, this survival time is shorter than for the no-planet case and we would expect to see $less$ grains beyond $r_{PB}$ than in this unperturbed case. But this depletion of very small fragments in the outer regions is partly compensated by the injection of slighly bigger collisional fragments due to planetary perturbations. This is clearly illustrated in Fig.\ref{cintf}, showing that $\beta \leq 0.12$ grains (with size $s\geq 4s_{cut}$) produced in the PB ring, which have an apoastron $\leq r_{PB}/(1-2\beta)$, and should in principle not be present in the $r>r_{PB}/(1-2\beta)$ region, contribute to up to $\sim$ 20\% of the brightness in these regions in the presence of a Jovian perturbing planet.

Profiles close to the reference -3.5 value are also obtained when considering planets on eccentric orbits (Fig.\ref{cintex}). In these cases, the stable ring of parent bodies is truncated much further inside the planet's orbit (for the $e_p=0.2$ case, it only extends to $\sim 31\,$AU and $\sim 37\,$AU in the periastron and apoastron directions, respectively)  but the profiles beyond this main ring are comprised between $\sim -3.4$ and $\sim -3.9$.

The only way to obtain a significantly steeper brightness profile is to assume a much lower collisional activity within the disc ($\tau_0=2\times 10^{-4}$). In this case, the balance between collisional injection of small grains and dynamical ejection by the planet is shifted in favour of the latter, and the slope of the luminosity profile is $\sim -5$.

\begin{figure*}

\makebox[\textwidth]{
\includegraphics[scale=0.4]{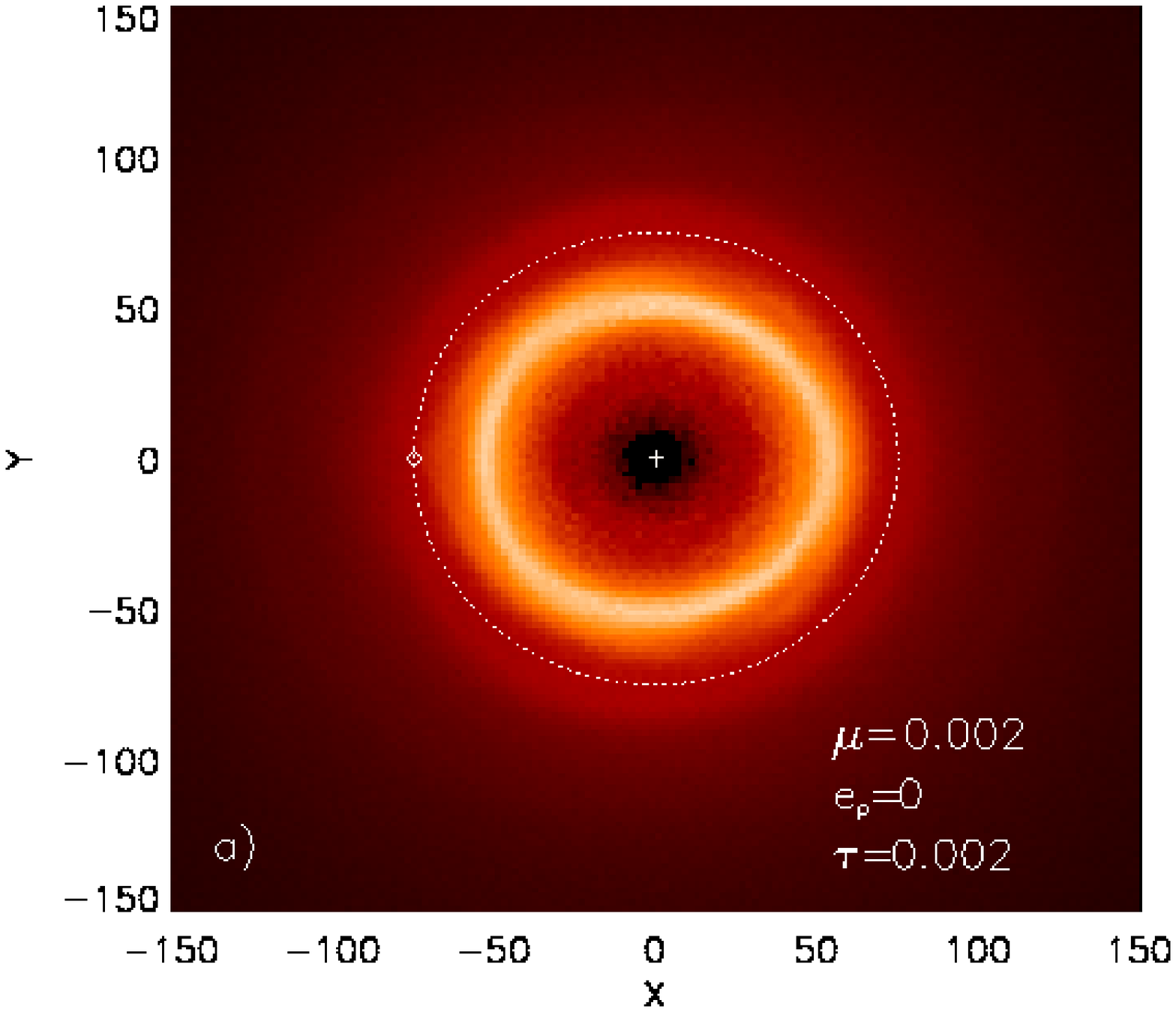}
\includegraphics[scale=0.4]{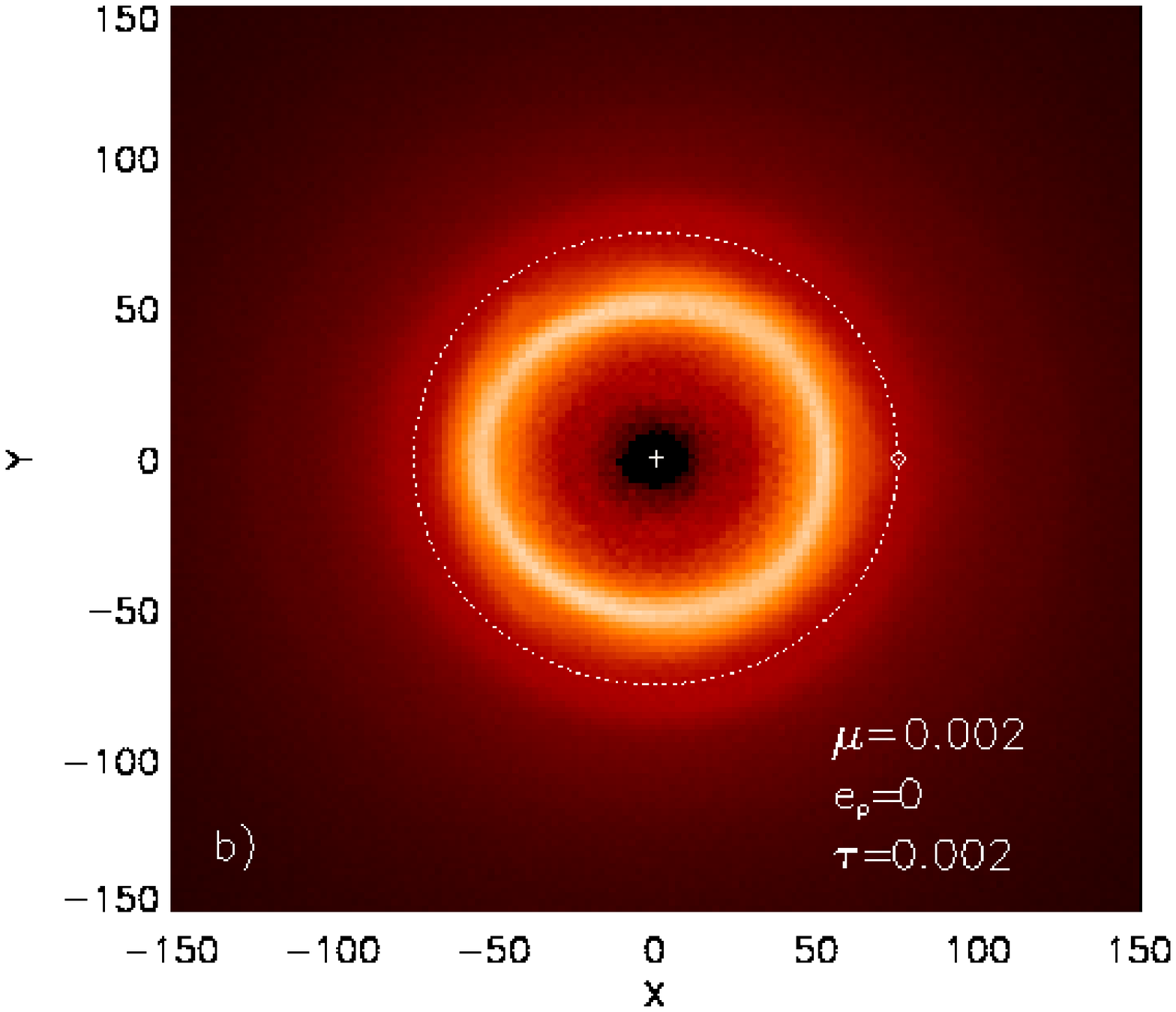}
}
\makebox[\textwidth]{
\includegraphics[scale=0.4]{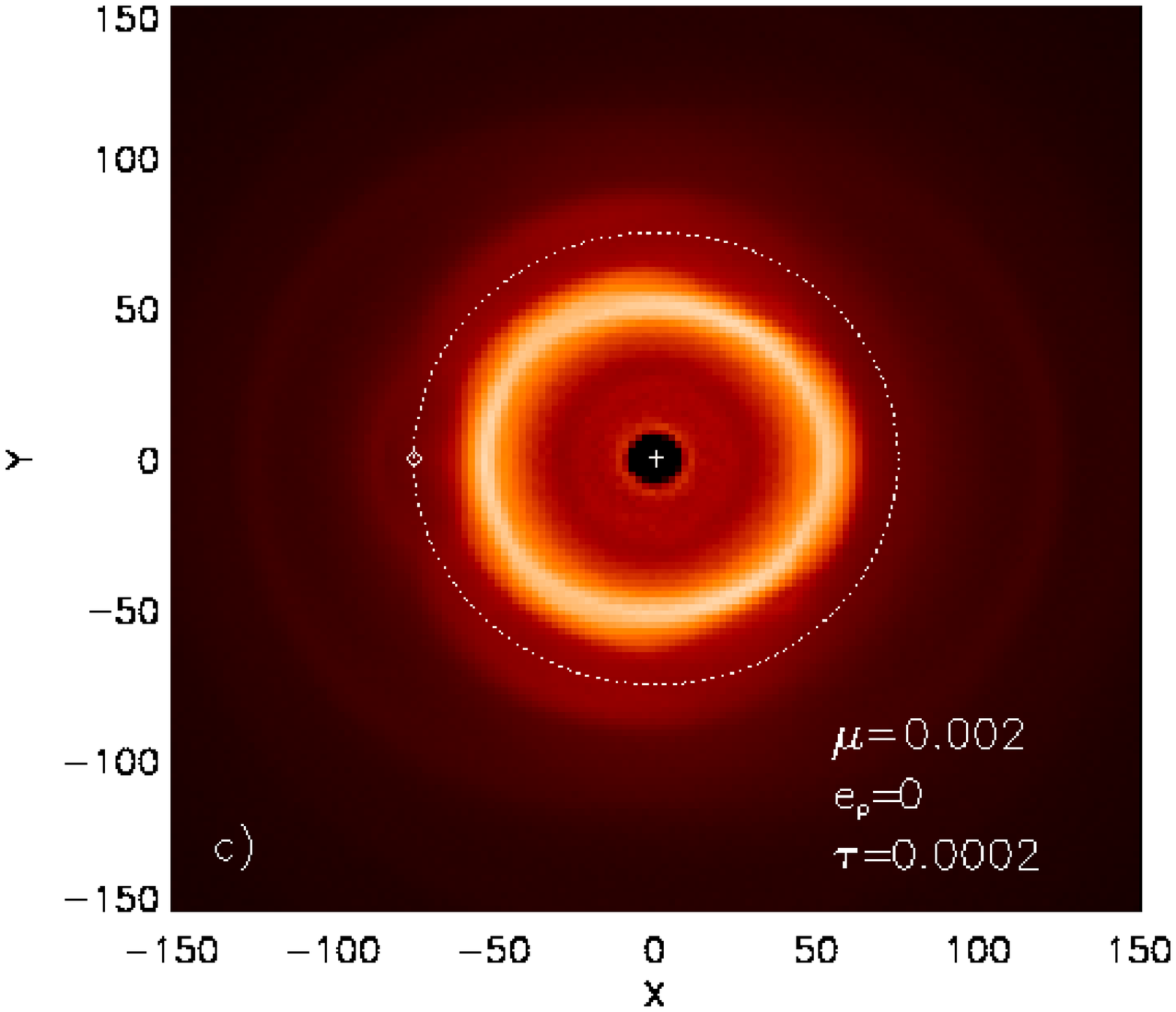}
\includegraphics[scale=0.4]{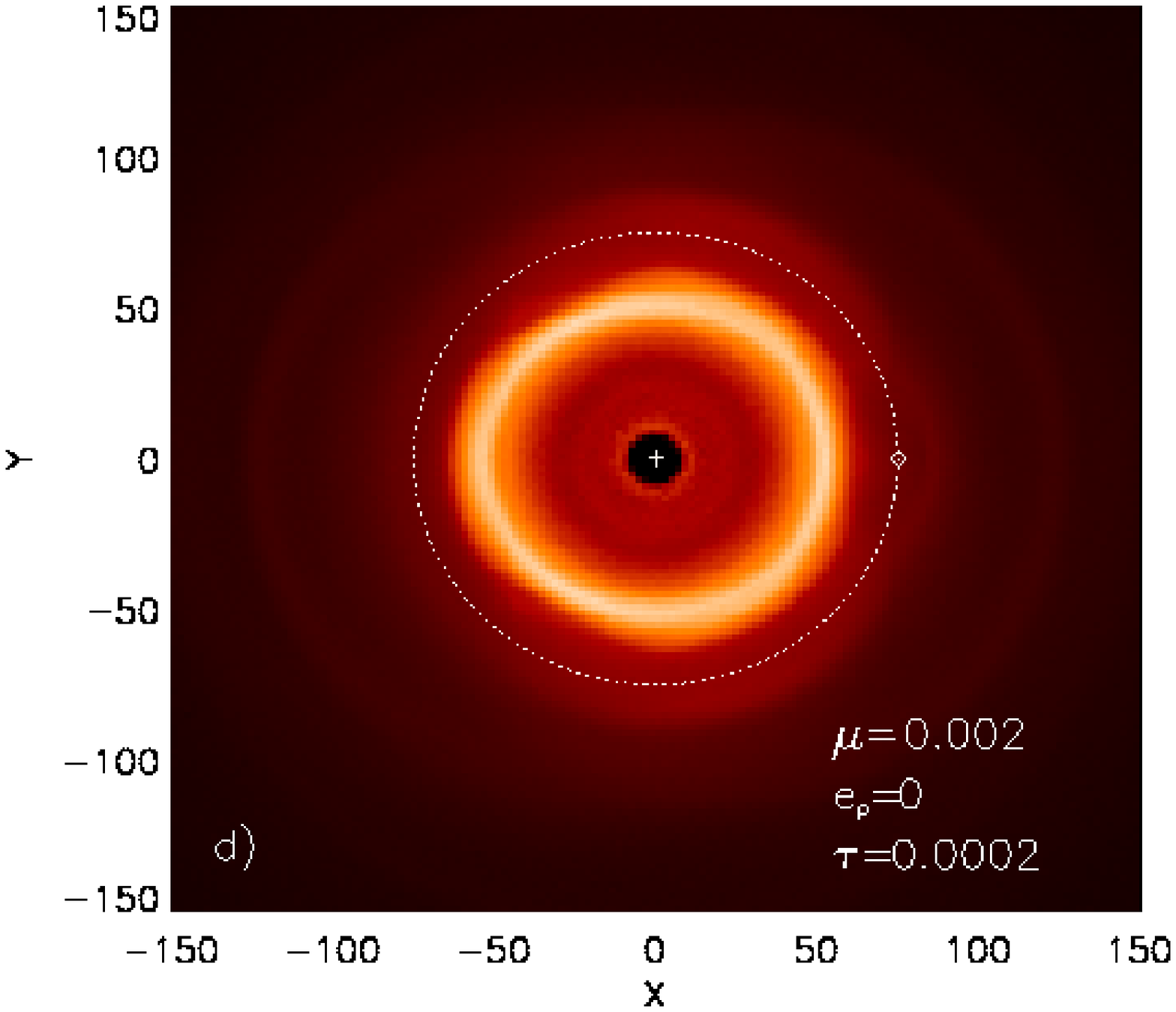}
}
\makebox[\textwidth]{
\includegraphics[scale=0.4]{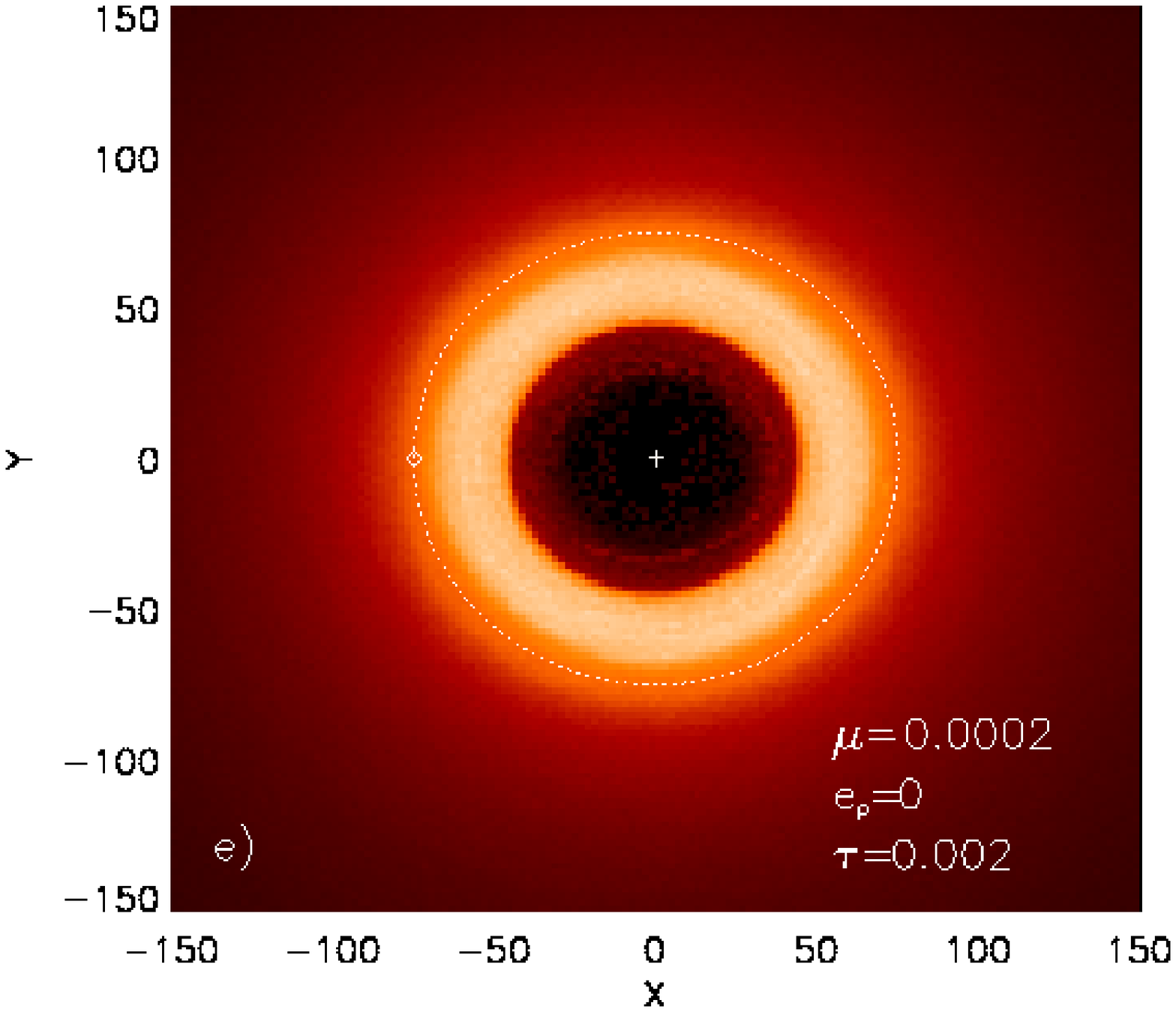}
\includegraphics[scale=0.4]{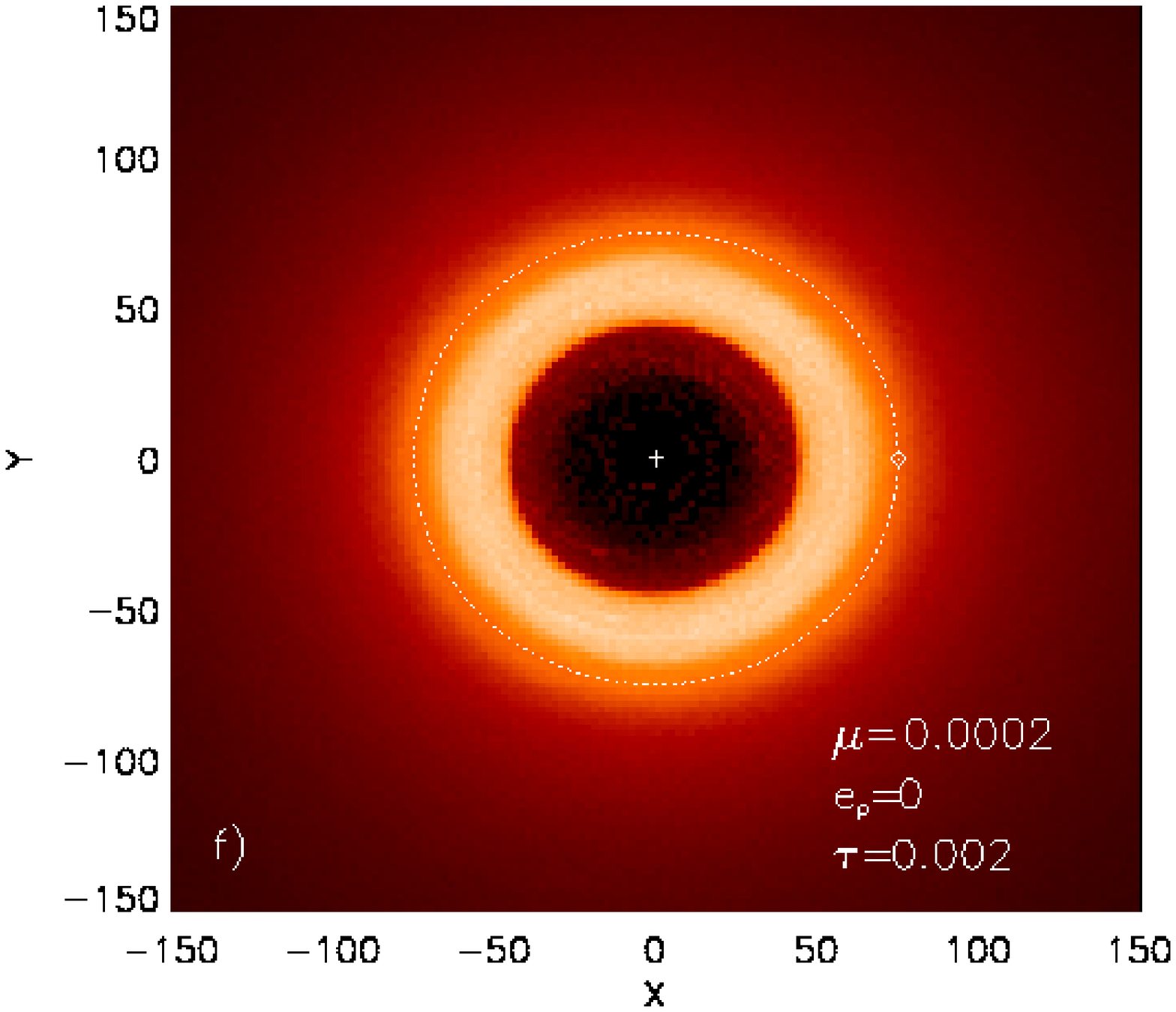}
}
\makebox[\textwidth]{
\includegraphics[scale=0.4]{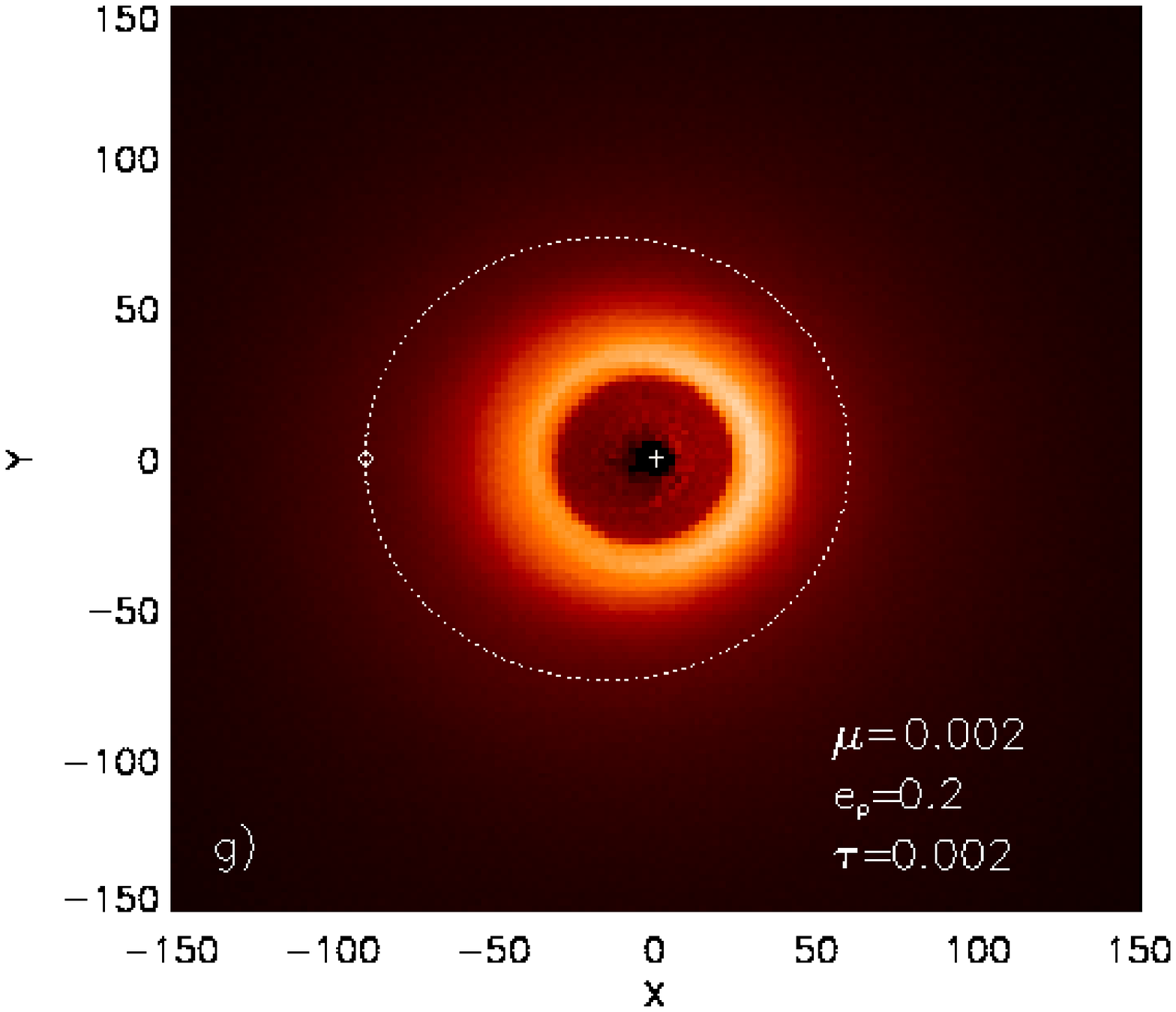}
\includegraphics[scale=0.4]{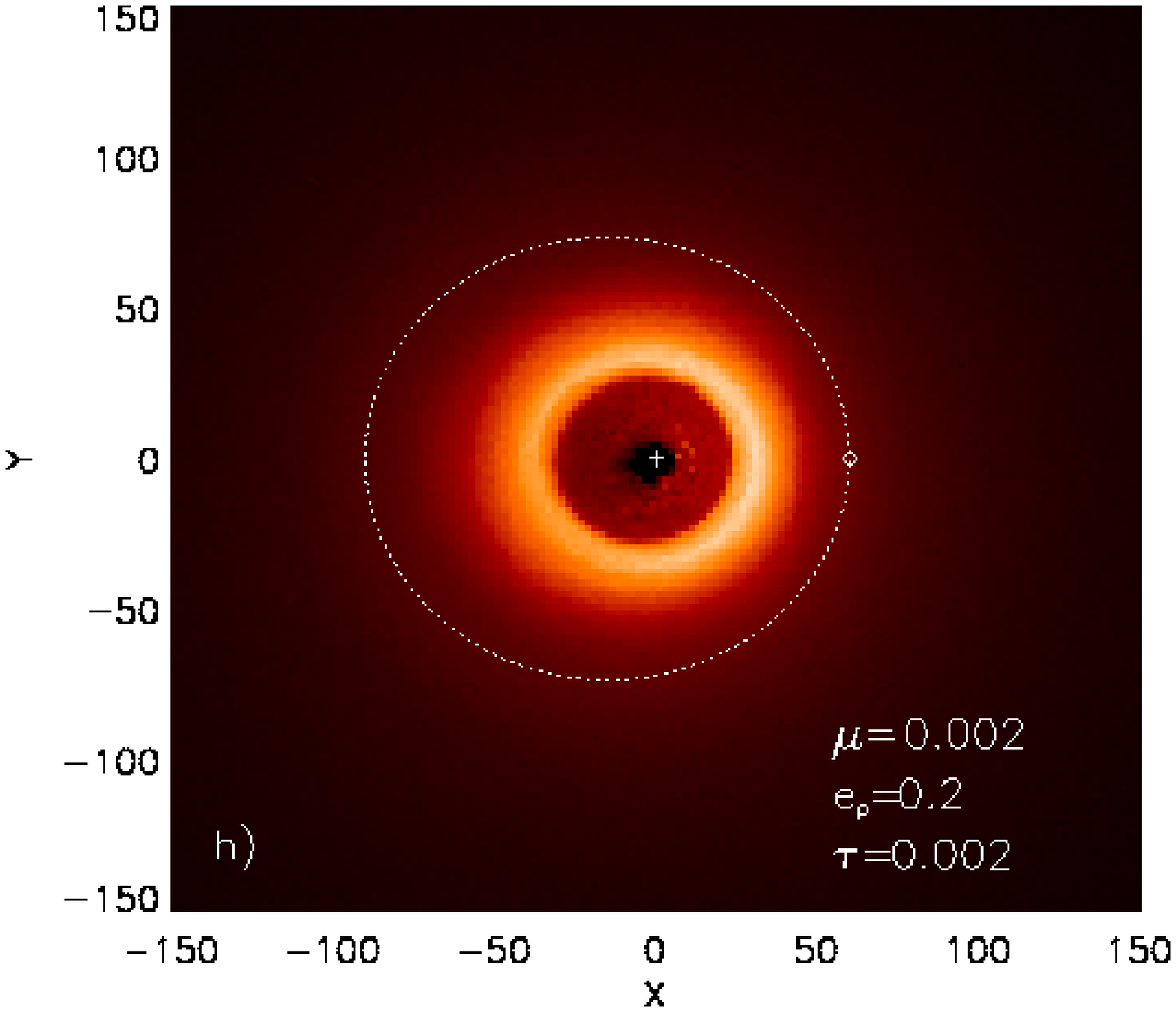}
}
\caption[]{Inner ring/outer planet: head-on synthetic images in scattered light, at steady state, for four different set ups. The graphs on the left hand side show the disc profile when the planet passes at apoastron, and the graphs on the right hand side correspond to periastron passages. The dotted curve shows the planetary orbit. An animated version of these graphs can be found at http://lesia.obspm.fr/perso/philippe-thebault/planpert.html
}
\label{collrunint}
\end{figure*}

When considering head-on images of the disc (Fig.\ref{collrunint}), it can be seen that, despite of the blurring effect of collisions coupled to radiation pressure, the planet's perturbations can induce significant spatial structures, mostly in the region of the parent body ring. For a $\mu=0.002$ perturber, the width of this main ring displays azimutal inhomogeneities, with three wider "blobs", located at $\sim \pm 30\,$ and $180$ degres of the planet's location, that precess with the planet. Not surprisingly, these strucutures are more prominent for the low collisional activity case ($\tau_0=0.0002$), but are still visible even in the nominal $\tau_0=0.002$ case. 
For an eccentric planet, these azimutal structures disapear, but the main ring takes on an eccentric shape aligned with the planet orbit's main axis. From the location of the peaks on the apoastron and periastron sides of the ring's radial profile (Fig.\ref{cintex}), we estimate the ring's eccentricity to be $e_{ring}\sim 0.08$ for $e_p=0.2$. This value is consistent with the forced eccentricity that can be derived from the secular theory of Laplace-Lagrange for this $\mu$ value and this disc-to-planet relative distance \citep[e.g.][]{murr99, must09}. Another clearly visible feature is the fact that the periastron side of the ring is brighter than the apoastron one. This is the pericentre glow effect identified by \citet{wyat99}, due to the fact that the periastron side of the ring, being closer to the star, receives and scatters more light. From Fig.\ref{cintex}, we see that the pericentre side is $\sim 1.35$ times brighter than the opposite one for the $e_p=0.2$ case, which compares well with the predicted difference in received light, i.e., $((1+e_{ring})/(1-e_{ring}))^2$ for the measured $e_{ring}=0.08$ value.

For a lower mass planet ($\mu=0.0002$), no significant spatial planet-induced structures are visible on the head-on images.

\section{Discussion} \label{Discu}

The main question the present study aims to address is if planets can significantly affect debris disc structures and more specifically if the presence of planets can be indirectly inferred from the spatial structure of an $observed$ disc. As noted in Section 2, we here implicitly consider observations in scattered light where the luminosity is dominated by the smallest grains in the disc, i.e., for wavelengths shortward of $\sim 5-8\,\mu$m. This is, however, not a too stringent limitation, since a large fraction of resolved images of debris discs, and in particular those with the best resolution, have been obtained in scattered light. The situation might, however, change with the upcoming high angular images expected from ALMA \citep[see][]{erte12}.

Another implicit assumption is that there is a minimum size $s_{cut}$ below which small grains are blown out by radiation pressure. This implies that the central star has a mass of at least $\sim 0.9M_{\odot}$. Note, however, that for lower mass stars of type M, stellar wind could play a role similar to that of radiation pressure around massive stars, blowing away grains below a given threshold size. This is for instance what is expected to take place in the AU Mic system \citep{auge06}.

Considering the already huge parameter space for the problem, all simulations were carried out for one given semi-major axis of the planet $a_p=75\,$AU. However, our results can be easily scaled to other planet separations with the same $relative$ disc/planet configuration, provided that the system's optical depth $\tau_0$ is the same. In this case, the collisional timescales follow the same scaling laws as the dynamical ones (those linked to the planet), i.e., $r^{1.5}$. The only difference between our reference case, with $a_p=75\,$AU and a $30 \leq r \leq 130\,$AU disc, and, say, a system with a $a_p=25\,$AU planet and a $10 \leq r \leq 43\,$AU disc is that all timescales should be shortened by a factor $\sim 5$. Note, however, that there is a limit to arbitrarily scaling up our results to high values of $a_p$, which is that impact $velocities$ in the disc should remain high enough to lead to erosive collisions. Apart from this global timescale effect, the three main parameters that control the system's evolution are the planet-to-star mass ratio $\mu$, its eccentricity $e_p$ and the disc's collisional activity, as controled by its average vertical optical depth $\tau_0$. All these parameters have been explored in our numerical investigation.

\subsection{Dectability of an embedded planet in scattered light images}

\subsubsection{Head-on viewed disc}

Our simulations have shown that the coupling of collisional activity and radiation pressure can efficiently attenuate or even erase some spatial structures that an embedded planet might create within a debris disc. This is particularly true in the feeding zone surrounding the planet's orbit. This region appears as a deep and well defined gap in the distribution of the parent body population, with the exception of the two Trojan populations at the Lagrangian points, but takes on a more homogeneous aspect once collisional production of small radiation-pressure-affected grains is turned on. 
This global tendency of collisions to attenuate spatial structures has also been witnessed by \citet{star09}, but for a different set-up, where the planet lies interior to a ring that migrates inwards because of Poynting-Robertson drag, and for smaller planet masses.
For the present case of an embedded planet, and even with massive, super-Jovian planets, azimutal structures like the Lagrangian points are no longer visible and the feeding zone appears as an almost continuous ring.
However, the $radial$ structure of the disc still bears the scars due to the presence of the planet: the feeding zone surrounding the planet's orbits is, for most cases, still less densely populated than the rest of the disc and appears as a gap-like structure. As a consequence, on 2-D face-on images, the presence of the planet, and even its radial location, can easily be inferred from this ring-gap-ring structure, and this even for Saturn-like or super-Earth planets. Although, in the latter case, the width of the gap, i.e., $\sim 0.05 a_p$ (Fig.\ref{collrun}c), might be too small to be observationally detectable.
In addition, for the specific case of a Jovian planet on a circular orbit (Fig.\ref{collrun}a), the structures due to the 2:1 resonance are still visible enough in order to also infer the azimutal position of the planet.
The only cases for which the planet does not leave a clearly identifiable gap in a head-on viewed collisional disc is if the planet has a high mass and is on an eccentric orbit. In this  case, the region exterior to the planet's location is strongly depleted and its contrast relative to the feeding zone is very low. However, in this case the planet's influence might be inferred from another feature, i.e., the eccentric shape of the $inner$ disc (Fig.\ref{collrun}d).

\subsubsection{Edge-on viewed disc} \label{edge}

For a system seen edge-on, the situation is much less favourable. Figs.\ref{cedgec}, \ref{cedgem} and \ref{cedgeex} clearly show that the signature left by the planet in the disc's radial luminosity profile is in most cases very weak. 

We firstly note that there is almost no asymmetry between the two opposite sides of the disc, i.e., the luminosity profile of the side where the planet is located is almost indistinguishable from the profile of the opposite side. Even for a perturber on a moderately eccentric orbit ($e_p\leq0.2$), the differences between the two ansae remain relatively limited, in any case too limited to be used as a smoking gun proof of the planet's presence.

Secondly, contrary to the parent body distribution, where there is a narrow and pronounced luminosity drop at the planet's location even for low mass planets (Fig.\ref{cedgenc}), in a collisional disc with small grains this sharp drop is replaced by a much wider, and often poorly defined and shallow hollow. For most explored parameters, the main effect of the planet is not this shallow hollow but a steepening of the luminosity profile in the radial region interior to the planet's position, as is best illustrated in the case of a $\mu=2\times 10^{-3}$ planet on a $e_p=0.2$ orbit (Fig.\ref{cedgeex}).
Given the spatial resolution obtained for the best resolved edge-on debris discs, such as $\beta\,$Pic \citep[e.g.,][]{goli06} or AU\,Mic \citep[e.g.,][]{met05}, this steepening should potentially be observable, at least for bright and nearby systems.
However, even if it is observed, it is doubtful that such a steeper profile could be unambiguously associated to the presence of a planet. Indeed, if on Fig.\ref{cedgem} the difference with the planet-less case is clearly visible, for real systems there would of course be no such "reference" case showing what the system should look like without a planet. Without this reference case, a steep slope could as well be associated to an intrinsic steeper radial distribution of the parent bodies (planetesimals) that could mimic the potential effect of a planet. 

In this respect, the small hollow that the planet induces in the profile would be a less ambiguous observable signature, provided it is pronounced enough to be observed, because it would be harder to mimic in a planetless disc where collisional production of small grains should naturally even out any short-scale variations in the density (i.e., luminosity) profile.
If we consider that a hollow is "pronounced" when there is an inflection point in the luminosity profile, then we see that this criteria is just about met for our nominal $\mu=2\times 10^{-3}$, $e_p=0$ and $\tau_0=2\times 10^{-3}$ case (Fig.\ref{cedgec}). Smaller planets (Saturn-mass and below) would induce much shallower hollows, as would also planets (even massive ones) on $e_p\geq0.1$ orbits. The same is true for denser discs ($\tau_0=0.01$) where the intense collisional activity eliminates the inflection in the radial luminosity profile. Conversely, the hollow is logically more pronounced for more tenuous discs ($\tau_0=4\times 10^{-4}$).
However, it remains to see if, even for the limited parameter space where the hollow is "pronounced", it would be concretely observable in real discs. For one of the best resolved edge-on discs, i.e., the one around AU\,Mic, such narrow dips seems indeed to be observed twice in the NW arm, once around $25\,$AU and once at $\sim 50\,$AU \citep[see Fig.4 of][]{met05}, with both times a depth that exceeds the plotted error bars. However, the presented error bars correspond to $1\sigma$ only. For more reliable $3\sigma$ error bars, none of the dips would remain visible. Furthermore, in both cases, these dips have no counterpart in the opposite SE arm, whereas our simulations predict that a planet-induced dip should be observable on both sides. Thus, we rate these structures as noise and conclude that the predicted inflection point is probably not observable directly with present instruments.

On the other hand, our simulations show that a clear \textit{jump} in the profile \citep[not just a knee as in][]{met05}, with a well defined, similar slope inside and outside the inflection point could reveal the presence of the inflection point without the need of detecting it directly (a local flattening of the profile as found for our $\tau_0 = 0.01$ case in Fig.\ref{cedgec} is already sufficient to produce the effect). Such a feature might be observable, because the slope of the surface brightness over a broad range of radii is usually well constrained as it is sampled by a large number of data points. Nevertheless, it is important to note that this is only a qualitative statement on the example of one observation. The exact sensitivity depends on a large number of parameters of the observation (e.g., integration time, PSF subtraction technique) and the disc (e.g., surface brightness, extent) and has to be evaluated from case to case. A full, quantitative discussion is beyond the scope of this work.
We note, however, that near-future instruments like \textit{VLT}/SPHERE \citep{doh06} are expected to significantly exceed present instruments in terms of sensitivity and contrast (i.e., the ability to observe fainter discs and/or at lower angular separation from the star). This will allow to obtain data of similar or increased quality of a larger sample of debris discs, improving the chances to detect the predicted structures.

Regardless of these considerations, it must be pointed out that, even if an inflection point is observable in the luminosity profile and if it can be linked to the presence of a planet, it does not directly give the radial position of the planet, since it is always located further out from the star. The offset between the planet's radial location and the position of the inflection point depends on the planet's mass and on the collisional activity in the disc. It varies between $\sim 0.05 a_p$ and $\sim 0.2 a_p$ for the parameters explored in our simulations (see Figs.\ref{cedgec} and \ref{cedgem}). Unfortunately, the value of the offset cannot be used to infer the planet's characteristics because the problem is degenerated: there is a trade-off between the planet's mass and the disc's collisional activity, which can both affect the amplitude of the offset.

\subsection{Ring sculpting by an outer planet}

\subsubsection{Can a planet truncate an inner ring?}

The dynamical sculpting of an exterior perturber has been often invoked as being a possible explanation for debris discs, or debris rings, with sharp outer edges, "sharp" meaning here sharper than the "natural" luminosity slope in $r^{-3.5}$ beyond the main ring's outer limit $r_{out}$ \citep{theb08}. 

\citet{theb10} and TBO12 have shown that a stellar companion can to some extent significantly deplete the region beyond $r_{out}$, even if it is $never$ able to fully truncate a collisionally active disc. The most effective depletion is obtained for a binary on a circular orbit lying as close as possible to the ring, i.e., so that the outer edge of the parent body ring corresponds to the limit beyond which orbits become unstable (in other words, the binary truncates the \emph{parent body} ring at $r=r_{out}$). In this case, the flux at a distance $2r_{out}$ from the primary is decreased by a factor 10 (with respect to the flux in the no-planet case) in the presence of a $\mu=0.5$ companion star on a circular orbit, and the slope can be as steep as $\sim r^{-8}$ in the $[r_{out},1.5r_{out}]$ region, before slowly decreasing again towards $r^{-1.5}$ at larger distances. For more eccentric binaries, the depletion effect is more limited, mainly because the companion spends much time at its apoastron far from the ring, thus leaving time for collisions in the ring to populate the "forbidden" $r>r_{out}$ domain \citep[see Fig.2 of][]{theb10}.

The present simulations show that planets are less effective than companion stars for truncating collisionally active debris rings. For a dense $\tau_0=2\times 10^{-3}$ ring, the slope of the luminosity profile beyond $r_{out}$ is relatively close to the -3.5 standard value in the planet-less case, and this regardless of the perturbing planet's mass. A maximum slope of only $\sim -4.2$ is reached for a $\mu=4\times 10^{-3}$ perturber (Fig.\ref{cint}). 

\begin{figure}
\includegraphics[width=\columnwidth]{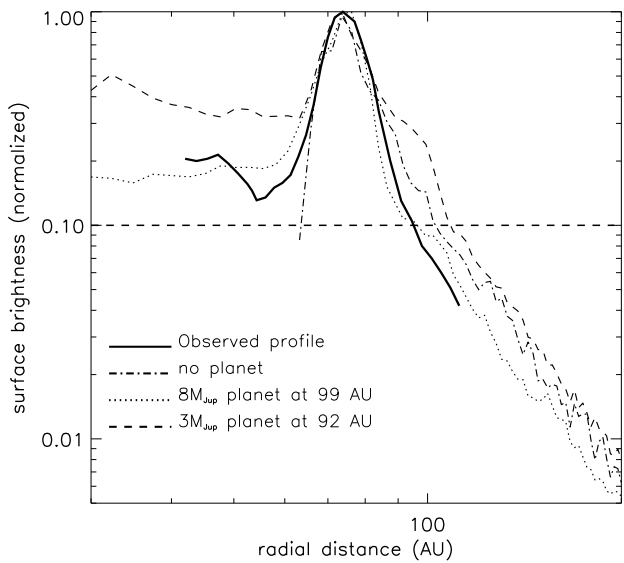}
\caption[]{Head-on viewed surface brightness profiles for the specific case of the HR4796A disc. The parent body ring is located between 67 and 75 AU. For each case, the planet is placed as close as possible to the parent body ring (i.e., so that the outer limit of the stability region coincides with the ring's outer edge). The solid line indicates the deprojected observed profile, derived from \citet{schn09}. The horizontal line delineates approximately the part (above this line) of the profile with a dynamical range of 10. As in \citet{lagr12}, we show, for each case, the radial cut at the "best" position angle, i.e., the one providing the best match to the observed profile.
Note that the goal of these runs is to fit the \emph{outer} profile, not the inner one, which should probably be shaped by other mechanisms \citep[see discussion in][]{lagr12}, so that not too much physical significance should be given to the profile interior to the ring's inner edge.}
\label{hr}
\end{figure}

Note, however, that these slopes are not reached immediately outside the parent body ring, but only after a small transition region of relative width $\Delta r/r_{out} \sim 0.2$, where the profile is steeper. This is in particular true in the planet-less case, where the slope of the luminosity profile is $\sim -7$ in the $[r_{out},1.2r_{out}]$ region. As a matter of fact, for small planets ($\mu \leq 2\times 10^{-4}$), the profile in this narrow region is $less$ steep than in the no-planet case, mainly because planetary perturbations inject medium-sized grains ($\beta \sim 0.1-0.2$) in these regions just beyond $r_{out}$ (Fig.\ref{cintf}). For more massive planets, the profile in this transition region gets steeper, but not much steeper than in the planet-less case.
This point is of importance for the specific case of the HR4796 ring, for which preliminary DyCoSS results were published in a recent paper \citep{lagr12}. The main conclusion of that paper, i.e., that the steep luminosity profile in $\sim r^{-9}$ outside the main ring could only be fitted with a massive outer planet has to be reinvestigated. Indeed, the region where the S/N of the observed profile is high enough does not extend very far outside the main ring (see Fig.16 of that paper), i.e., not far outside the narrow transition region where slopes could be high even for a planet-less case, a case that was not considered in \citet{lagr12}. We present in Fig.\ref{hr} a revised version of that paper's Fig.16, taking this time into account the no-planet case. As can be seen, in the 75-90 AU region where the observed profile has a reliable S/N ratio, the no-planet case does in fact give a better fit than the "small" planet case ($3 M_{Jup}$). However, the massive planet ($8 M_{Jup}$) case\footnote{Note that, since HR4796A is a massive A0V star of mass $2.2\,M_{\odot}$, the 8 and $3 M_{Jup}$ cases correspond to $\mu = 0.0035$ and $\mu=0.0013$, respectively} is still the one giving the best fit, so that the main conclusion of \citet{lagr12} still holds. But to unambiguously discriminate between the different scenarios, reliable observations further outside the main ring are clearly necessary.

As a matter of fact, there is to our knowledge no disc for which a steep ($\leq r^{-5}$) luminosity profile has been observed far outside the expected location of the main parent body ring \footnote{We here define the outer limit of the main ring as the outer limit of the large \emph{parent bodies} ring, which roughly corresponds to the point where the scattered light luminosity peaks. This outer limit is thus different from the one usually given in observational studies, i.e., the radial distance where the flux falls to half that of the peak. From our study we find that this observational outer limit is actually located beyond the outer edge of the main collisionally active ring. For most cases, this limit is close to the outer edge of the transition region, with steeper slopes, just outside the main PB ring}. For the two archetypal examples of rings with steep luminosity decreases, Fomalhaut and HR4796, these profiles have not been reliably obtained beyond the expected transition region just outside the main ring. 

From our simulations, the only way to get profiles that are significantly steeper than in the planet-less case, both in the narrow transition region outside the PB ring and further out, is when assuming a much lower collisional activity in the disc. This is clearly illustrated by the $\tau_0=2\times 10^{-4}$ run in Fig.\ref{cint}, for which the slope in the transition region is $\sim -10$ and tends towards $\sim -5$ further out. Such a low $\sim 2\times 10^{-4}$ optical depth is, however, much lower than the one derived for HR4796, $\sim 5 \times 10^{-3}$ \citep{auge99,wyat99}.
Nevertheless, $\tau_0 \sim 2\times 10^{-4}$ values are comparable to those derived for other resolved debris rings, in particular Fomalhaut \citep{bole12} and HD202628 \citep{kris12}. The $\sim -5$ value we find could be, given the uncertainties on the data, compatible with the measured slopes in the outer regions of these 2 systems, i.e., $\sim -4.6$ to $\sim-6.5$  for Fomalhaut\footnote{The value given in \citet{kala05} is -4.6, however, from the data given in Fig.3 of that paper we find a best fit value of $\sim -6.5$} \citep{kala05} and $\sim -4.7$ for HD202628. However, the radial extent of the reliable data for Fomalhaut does not exceed $\sim 1.15 r_{out}$ and is thus confined to the "transition" region mentioned before, so that no conclusions can be drawn for this system. For HD202628, on the contrary, it reaches far outside this transition region, i.e., $\sim 1.5 r_{out}$, so that this system could in principle be explained by the presence of an outer jovian planet. However, given that the radial luminosity of this systems peaks at $\sim 190\,$AU, the planet would have to be located at $\sim 280$\,AU (as implied by rescaling Fig.\ref{cint}), which is much further out than any exoplanet detected to date.

\subsubsection{Ring morphology}

Another important issue is the spatial structure of the main parent body ring itself. Our simulations show that massive planets on circular orbits can induce azimutal inhomogeneities that precess with the planet. These inhomogeneities remain visible in scattered light even for a high collisional activity within the ring, albeit with lower contrast (Fig.\ref{collrunint}). 

As of today, no such intrinsic brightness inhomogeneities have been unambiguously observed in the 7 ring-like discs that have been imaged in scattered light \citep[see Section 4.6 of][]{kris12}, apart from those that might depend on scattering phase angle. Some bright spots have been observed in the HR4796A ring, but due to the limited signal to noise of the data, it is possible that it could be an artifact \citep{thal11,lagr12}. 
Note also that the detection of such precessing inhomogeneities would not be a smoking-gun proof of the presence of an outer planet, as an \emph{inner} planet would also produce such precessing structures. There is, however, a way to discriminate between these two hypothesis, by measuring the precession rate of the structures: if it is faster than the orbital period at the ring's distance, then it is due to an inner planet, if it is slower then it is due to an outer perturber. Depending on the quality of the precession measure, the orbital period and thus the semi-major axis of the planet can in principle be estimated.

For planets on eccentric orbits, no precessing azimutal structures are visible in our results, but the main ring becomes itself eccentric, with $e_{ring}$ close to the secular forced eccentricity. We observe the well known pericentre glow effect, i.e., a brightness asymmetry between the periastron and apoastron sides of the ring. Our simulations show that this effect remains observable, in scattered light, even in collisionally active systems. The formation of an eccentric ring in response to an outer planet's perturbations is an interesting outcome, as several such eccentric rings have been observed, in particular around Fomalhaut and HD202628 \citep{kris12}. We stress, however, that the presence of an outer planet \emph{alone} cannot explain all the characteristics of these rings. This is particularly true for the inner edge of these rings, which is always very sharp and has to be sculpted by "something", most probably an \emph{inner} planet \citep[e.g.][]{chia09}. If this inner planet is on an eccentric orbit, then the ring naturally assumes an eccentric shape without the need of an additional external planet. The question is then if there are other ring characteristics that cannot be explained without invoking such a planet. Of course, an outer perturber could easily explain the sharp outer edge of the \emph{parent body} ring. Such a sharp edge is indeed a prerequisite to obtain a scattered light brightness profile decreasing as $r^{-3.5}$, or even steeper, beyond it. For Fomalhaut, recent 850\,$\mu$m ALMA images \citep{bole12} provide unprecedented information about the spatial distribution of the large parent bodies and show that they indeed form a very confined ring.
\cite{bole12} argue that such a confinement can be produced by two Earth-to-super-Earth shepperding planets. There is however a potential problem with this hypothesis when considering the brightness profile in scattered light. Indeed, our simulations show that, even for the low $\tau_0 \sim 10^{-4}$ optical depth estimated for Fomalhaut, an Earth-mass planet cannot prevent the brightness profile from being $\sim -3.5$ beyond the main ring. This seems to be in contradiction with the steep slopes, between -4.6 and -6.5, that have been observed in scattered light images. However, let us again stress that these profiles have not been reliably obtained beyond $\sim 160$\,AU, less than 20\,AU beyond the main ring's outer limit at $\sim 140$\,AU, which is still inside the transition $\sim [r_{out},1.2r_{out}]$  region where steep slopes can be obtained regarless of the presence of a planet (Fig.\ref{cint}). Additional scattered light observations, probing the Fomalhaut disc further out from the main ring, are needed in order to confirm or invalidate the outer-planet hypothesis. 

An alternative explanation for a sharp outer edge of the parent body ring could be that this is the natural outer limit beyond which planetesimals could not form during the proto-planetary disc phase. Our simulations show that it would be difficult to discrimate between such a system, with an inner eccentric planet close to the "natural" outer limit of a debris disc, and a system with the same inner planet and an outer planet on an eccentric orbit truncating the disc.
We note that an even more radical alternative scenario, dispensing even with the inner planet, has been recently proposed to explain the formation of sharp and eccentric debris rings in a planetless environment \citep{lyra12}. This issue is thus still an open one.

\subsection{Comparison with previous works and perspectives}

As mentioned in Sec.1, the present study is by no means the first numerical investigation of the signature left by a planet in a debris disc. It is, however, one of the first to take into account the crucial role played by collisions within the disc, whereas most past studies were based on collisionless dynamical simulations of size-less particles \citep[e.g.,][]{kuch03,rech08}, or, in more advanced versions, a combination of such collisionless simulations for single-sized particles that were later recombined assuming a fiducial "collisional equilibrium" size distribution for all disc particles \citep{quil02,wyat06,erte12}.

As such, it is not surprising that our results sometimes significantly depart from those obtained in pure N-body studies. 
As an example, we compare our results to those obtained by the most recent of such studies, that of \citet{erte12}, which produced synthetic images of planet-bearing discs from visible to far-IR and millimetre wavelengths. As mentioned earlier, at such long wavelengths the flux is no longer dominated by the smallest grains and our modelling is not valid. Thus, we only compare our results with their ones up to $\lambda=10\mu$m. One important conclusion of \citet{erte12} was that Jovian planets always induced prominent structures in discs in scattered light, and that such structures might in most cases be observable with near-future facilities, at least in face-on oriented discs. When we compare our disc structures (Fig.\ref{collrun}) with the ones shown in \citet{erte12} for the face-on case, we find that the resonant structures found there to appear due to transport mechanisms (Poynting-Robertson drag and stellar wind drag) disappear when considering collisions. In contrast, we observe a horseshoe structure and a clear gap in the disc, which were observed by \citet{erte12}, but only at longer wavelengths and not in scattered light. This is because the absence of collisional lifetimes in \citet{erte12} allowed Poynting-Robertson drift to populate the gap around the planet's location even with large grains. We do not observe this inward drift in the present simulations, even for the lowest collisional activity considered, because the  collision destruction rate of large grains is always higher than typical timescale for PR drag. As a matter of fact, the distribution of the biggest grains considered in our runs, of size $s=40s_{cut}$, is undistinguishable from that of the parent body simulations.
In the present runs, the grains that are present in the resonant and horse-shoe structures were \emph{not} captured when drifting inward from an outer region, but were present in these regions from the beginning. 
 
\citet{erte12} also obtained spatial structures for edge-on seen discs, albeit less pronounced than for head-on systems. Our simulations have shown that such structures can probably not remain in a collisionally active system.

The only other debris disc model that incorporates a degree of collisional processes into an N-body scheme is the CGA of \citet{star09}.  
It is, however, difficult to compare results obtained with both codes since they have not been designed to address the same perturber/disc configurations and have so far been used for different set-ups. CGA has been used to investigate the case of a perturbing planet interior to an outer ring, where an important effect is the slow drift due to PR-drag from ring particles into the planetary region. This drift results in efficient trapping into outer mean motion resonances and thus azimutal structures much more pronounced than for non-drifting particles\footnote{\citet{kuch10} results regarding the respective balance between PR drag and collisions are more complex than this simplified statement, and depends on the value of the system's average $\tau$ as compared to a critical "crossover" optical depth $\tau_r$}. On the contrary, PR drag has a negligible effect for the two set-ups that have been considered in the present study, i.e., embedded planet and inner-ring/outer-planet. In the former, collisional timescales within the disc are always much shorter than PR drag ones. In the latter, PR drag is observable, but its main effect is to progressively move small grains into the region interior to the ring's inner edge (see for instance Fig.\ref{hr}), i.e., far from the region of interest between the planet and the ring's $outer$ edge.
Note, however, that our results for the embedded-planet case regarding the insignificance of inward drift of big grains towards the planet is consistent with that of \citet{kuch10}, who did not observe any drift of large particles from an outer belt towards their inner planet. In both cases, finite collisional lifetimes do prevent this drift from happening. The apparent discrepancy with \citet{kuch10}'s synthetic images (see their Fig.9), i.e., the presence of big grains trapped into horse-shoe structures and MMRs, simply follows from different initial set-ups: a disc-\emph{embedded} planet in the present runs, as opposed to a planet lying in an empty region well interior to an exterior belt.
In addition to this difference in the planet-to-disc configurations, DyCoSS and CGA have also considered different planet masses. \citet{star09} and \citet{kuch10} have explored perturber masses in the Earth-to-Neptune mass range, while TBO12 and the present study mostly focused on stellar perturbers and massive, Saturn to super-Jovian planets.

Finally, we want to underline that DyCoSS (as CGA) is \emph{not} a fully integrated dynamics+collisions model. Although some important issues can be investigated with it, in particular those related to timescales and production rates, the treatment of collisions relies on simplifying assumptions and is not fully incorporated into the scheme of the code. Furthermore, only systems that have reached dynamical and collisional steady state can be studied. It is not suited to investigate transient or violent events that might take place in many debris discs.
As emphasized in the Introduction, fully-integrated codes are still out of reach today but improvement in computational capacities should hopefully make them available in a close future.

\section{Conclusions}

Using the DyCoSS code, we numerically investigate the effect of collisions on the formation and survival of planet-induced structures in debris discs.
We confirm the qualitative result found by some previous studies using less sophisticated models, i.e., that collisions have a global tendency to even out sharp dynamical structures. We are here, however, for the first time\footnote{With the notable exception of the collisional grooming algorithm of \citet{star09}, which is, however, intended for a different set-up and has been used for a different range of planet masses} able to take this general result a step further by quantifying this effect and its consequences on observability.

For a planet embedded in an extended disc, we find that the planet's signature in the disc could probably remain visible if the system is seen head-on, provided that the planet is either on a nearly circular orbit and has a mass $\geq 2\times 10^{-4}M_\star$, or that it has a mass $\geq 2\times 10^{-3}M_\star$ if it has a $e_p\sim 0.2$ orbit. If the system is seen edge-on, the situation is much less favourable: for bright and collisionally active discs, collisions are able to erase most small-scale signatures induced by the presence of a planet, although a small hollow in the luminosity profile might be detectable under very favourable conditions. Large-scale structures, in particular a steepening of the radial luminosity profiles interior to the planet's orbit, should in principle be detectable, but they might be very difficult to unambiguously associate to the presence of a planet, as other causes could mimic this effect.

For the other configuration we considered, a planet exterior to an inner ring, we find that the planet is never able to prevent collisions and radiation pressure from populating the regions beyond the main ring with small fragments, even when these regions are dynamically unstable. In fact, the luminosity profile beyond the main ring is in most cases relatively close to what it should be in a planet-less case, i.e., decreasing as $r^{-3.5}$. Only for tenuous discs with optical depths in the $\leq 2\times 10^{-4}$ range do we obtain significantly steeper profiles. 
There is, however, a narrow transition region just outside the main ring where steep luminosity profiles can be obtained, but here again the presence of a planet does only moderately affect their slope.
The planet can, however, affect the morphology of the region corresponding to the main parent body ring. In this region, the signatures of a planet on a circular orbit are azimutal inhomogeneities that precess with the planet. These features remain visible despite the blurring due to the coupling of collisions and radiation pressure. Detecting and measuring the precessing rate of such features in observed discs could be an unambiguous way to indirectly infer the presence of an outer planet, and even constrain its location.
If the planet is eccentric, then its signature is to render the main ring elliptic and to induce the well known pericentre glow effect (periastron side brighter than the apoastron side), which is still clearly visible even in highly collisionally active systems. However, such features, if detected in real discs, cannot be unambiguously attributed to a putative outer planet, as other causes, such as an inner planet close to the "natural" outer limit of the planetesimal disc, could induce similar signatures.

For both considered set-ups, lowering the collisional activity in the disc to values of $\tau_0$ of a few $10^{-4}$ results in more pronounced spatial structures. In other words, for a given planet mass and orbit, there is a trade-off between the brightness of the disc and the sharpness of planet-induced structures: the brighter the disc, the higher the collisional activity and thus the lower the sharpness of the spatial features.

DyCoSS, together with the CGA of \citet{star09}, is a first step towards the next generation of debris discs models, where collisions and dynamics will be self-consistently integrated into the model's structure.

\begin{acknowledgements}

Q.K. and S.E. are funded by the French National Research Agency (ANR) through contract ANR-2010 BLAN-0505-01 (EXOZODI). P.T. ackowledges financial support by the same contract.

\end{acknowledgements}

{}
\clearpage

\end{document}